\documentclass[11pt]{article}
\usepackage{graphicx}

\usepackage[final]{acl}

\usepackage{times}
\usepackage{latexsym}
\usepackage[utf8]{inputenc}
\usepackage{kotex}
\usepackage{pgfplots}
\usepackage{longtable}
\pgfplotsset{compat=1.18}

\usepackage[T1]{fontenc}

\usepackage[utf8]{inputenc}

\usepackage{microtype}

\usepackage{inconsolata}

\usepackage{booktabs}
\usepackage{multirow}
\usepackage{amsmath}
\usepackage{amssymb}
\usepackage{xcolor}
\usepackage{float}
\usepackage{tcolorbox}
\usepackage{enumitem}
\tcbuselibrary{skins, breakable}
\usepackage{adjustbox}
\usepackage{cuted}
\tcbset{
  promptbox/.style={
    enhanced, breakable,
    colback=gray!6,
    colframe=gray!40,
    boxrule=0.6pt,
    arc=4pt,
    left=8pt, right=8pt, top=6pt, bottom=6pt,
    fonttitle=\bfseries\small,
    attach boxed title to top left={yshift=-2mm, xshift=6pt},
    boxed title style={
      colback=gray!30,
      colframe=gray!40,
      arc=3pt,
      boxrule=0.5pt,
    },
  },
  systemprompt/.style={
    promptbox,
    colback=blue!4,
    colframe=blue!30,
    boxed title style={colback=blue!20, colframe=blue!30, arc=3pt, boxrule=0.5pt},
  },
  userprompt/.style={
    promptbox,
    colback=green!4,
    colframe=green!30!gray,
    boxed title style={colback=green!15, colframe=green!30!gray, arc=3pt, boxrule=0.5pt},
  },
  outputschema/.style={
    promptbox,
    colback=orange!4,
    colframe=orange!40,
    boxed title style={colback=orange!20, colframe=orange!40, arc=3pt, boxrule=0.5pt},
  },
}

\title{CORA: A Protocol for Diagnosing Boundary Robustness in Text-to-Audio Retrieval under Query Reformulations}

\author{
  \textbf{Jae Min Woo}\thanks{These authors contributed equally to this work.},
  \textbf{Kyongmin Kong}\footnotemark[1],
  \textbf{Bogyung Jeong}\footnotemark[1],
  \textbf{Minjeong Kim}\footnotemark[1]\\
  \textbf{HaeJun Yoo},
  \textbf{Du-Seong Chang}\thanks{Corresponding author.}\\
  Sogang University\\
  \small{\texttt{\{wjm9765, kongmin00, jbg9277, essibae5, judejiwoo, dschang\}@sogang.ac.kr}}
}

\begin{document}
\maketitle

\begin{abstract}
Text-to-Audio (T2A) retrievers are typically evaluated with caption style queries, but the same user intent can be expressed in many forms.
We introduce CORA (Caption-Offset Retrieval for Audio), a caption anchored diagnostic protocol that rewrites each source caption into five intent preserving forms (Command, Question, Indirect, Key phrase, and Statement) while fixing the target audio.
By tracking the same target across query forms, CORA defines RankDrop, a metric revealing failures hidden by Recall@k.
Using Pearson's correlation coefficient $r$, we find that RankDrop is weakly associated with raw text space movement ($r=0.084$), but strongly associated with Target Alignment Loss and Target Boundary Margin Degradation ($r=0.508$ and $r=0.615$).
The same pattern appears in OEA retrievers, where RankDrop is better explained by boundary degradation ($r=0.472/0.478$) than by query movement ($r=0.084/0.046$).
Overall, these results suggest that robust T2A retrieval requires preserving the target's boundary advantage over competing audio under reformulation.\footnote{We release our code and the CORA dataset at \url{https://github.com/sogang-isds/CORA}.}

\end{abstract}

\section{Introduction}

Text-to-audio (T2A) retrieval systems are typically evaluated with caption style text queries.
However, the same semantic intent can be expressed in substantially different linguistic forms including imperative command, a yes/no question, a polite indirect request, a keyword phrase, or a declarative statement.
A robust retriever should preserve the target audio across these intent-preserving reformulations, but standard Recall@k evaluations provide only aggregate performance and cannot reveal whether the same target experiences a rank decline when the query form changes.

Query variation can substantially affect retrieval robustness in text information retrieval~\citep{penha2022robustness}, and real sound search queries differ markedly from the caption style descriptions commonly used in T2A benchmarks~\citep{weck2024language}. 
RobustCLAP addresses linguistic variation through multi-view contrastive learning~\citep{selvakumar2025robustclap}, while OEA introduces user intent queries and Hard Negative oriented evaluation for audio-text retrieval~\citep{oea2025}. 
These studies show that query form robustness matters and that dedicated training or evaluation can improve it. 
In this work, we ask a complementary diagnostic question: when retrieval failures persist despite intent preserving reformulation, what embedding space change drives them?

We introduce CORA, a diagnostic protocol anchored to source captions
for T2A retrieval under expressive query reformulations.
Given a benchmark caption, CORA generates five semantically aligned query forms (Command, Question, Indirect, Key phrase, and Statement) while keeping the target audio and candidate pool fixed.
CORA tracks the rank of the same target audio before and after reformulation, revealing that reformulation effects depend on the model–dataset regime: a query form that is stable in one setting can be harmful in another.

We then analyze the mechanisms behind these rank drops. We compare a text space movement account with a target boundary account, which examines the target's score advantage over competing non target audio. Across 66,100 CORA pairs, raw query movement is only weakly associated with RankDrop ($r=0.084$). Conversely, Target Alignment Loss (TAL) and Target Boundary Margin Degradation (TBMD) show much stronger associations ($r=0.508$ and $r=0.615$). Notably, this boundary degradation pattern persists in external robustness oriented retrievers like OEA and RobustCLAP, indicating that preserving the target margin is a core aspect of stable retrieval.

Our contributions are as follows:
\begin{itemize}
    \item We introduce CORA, a caption anchored diagnostic protocol for T2A retrieval that fixes the target audio and candidate pool across five intent preserving reformulations.
    
    \item We show that query form sensitivity is not determined by surface form alone, but depends heavily on the model and dataset regime.
    
    \item We provide paired rank diagnostics demonstrating that residual retrieval degradation(even in robustness oriented retrievers) is better characterized by target alignment loss and boundary degradation than by raw text space movement.
\end{itemize}

\section{Related Work}
\subsection{Audio-Language Models}

Contrastive Language-Audio Pretraining (CLAP) has become the dominant paradigm for T2A retrieval, learning joint embedding spaces via InfoNCE loss \citep{wu2023laionclap, elizalde2023clap}.
Scaling along data and model axes has yielded consistent gains \citep{mei2024wavcaps, msclap}, while a complementary line of work addresses structural failures within the standard alignment space: \citet{ghosh2024compa} exposed limitations in compositional reasoning; \citet{li2024mgaclap} introduced locality-aware features to recover fine-grained correspondences; \citet{niizumi2025m2dclap} combined masked modeling with contrastive pretraining; and \citet{yuan2024tclap} proposed temporal-contrastive captions to recover ordering information.
These contributions illuminate representational limits of CLAP, but leave open why retrieval degrades under linguistic variations of the same semantic intent --- the question CORA is designed to answer.

\subsection{Query Robustness in Audio and Text Retrieval}
Query brittleness is well-documented in information retrieval. For instance, \citet{penha2022robustness} showed that neural pipelines suffer roughly a 20\% drop in effectiveness under meaning-preserving paraphrases. More recently, \citet{weck2024language} identified a stark distributional gap between real Freesound user queries (averaging just 1.8 tokens) and the lengthy, multi-word captions conventionally used for training and evaluation \citep{kim2019audiocaps, drossos2020clotho}. While \citet{selvakumar2025robustclap} mitigated paraphrase-driven performance drops via multi-view contrastive fine-tuning, and \citet{oea2025} exposed hidden retrieval limitations using a five-type User-Intent Query (UIQ) taxonomy, these approaches largely focus on aggregate robustification or macro-level evaluation.
In contrast, our work adopts a granular, instance-level perspective, tracking whether a specific target remains retrievable under systematic query form shifts. Rather than proposing new training objectives or architectural modifications, we focus on a diagnostic objective: characterizing residual failures through directional geometry in the joint embedding space—a subtle failure signal that aggregate metrics inherently fail to capture.
\subsection{Contrastive Embedding Geometry and Hard Negatives}

Hard negative mining is foundational to contrastive learning \citep{robinson2021contrastive}, and has been applied to improve compositional reasoning, fine-grained alignment, and text retrieval at scale.
\citet{jiang2024hardnegative} show that under hard negative sampling, optimal representations converge to neural-collapse configurations, making embedding geometry a precise object of study.
In vision-language models, \citet{liang2022modality} identified a modality gap in CLIP arising from initialization and contrastive loss dynamics; \citet{kang2025clip} proved formally that no CLIP-like space can simultaneously achieve correct attribute binding and compositional understanding.
We deploy hard negative as \textit{inference-time diagnostic probes}, measuring whether a query-shift vector moves toward the target and away from the decision boundary between the target and its hard negatives. 
This signal captures precisely the mechanism by which form shifts displace a target behind its nearest competitor.

\section{Method}
\subsection{Caption-Offset Retrieval for Audio (CORA)}
Most existing audio-text retrieval benchmarks rely on caption-style queries, yet real users frequently express the same intent as commands, questions, indirect requests, keyword phrases or declarative statements. To evaluate robustness to diverse user formulations, we define five query types commonly observed in real-world audio retrieval scenarios: Command, Question, Indirect, Key phrase, and Statement queries.

Our taxonomy is based on the query formulations introduced in UIQ~\citep{oea2025}, but we modify several query types and generation strategies to better reflect the diversity of real-world user expressions (Section~\ref{sec:cora_generation}). CORA uses this taxonomy to evaluate retrieval robustness. Sections~\ref{sec:metrics}--\ref{sec:projection_diagnostics} define the paired evaluation and the embedding-space diagnostics that make up the actual protocol. Full query definitions and examples are provided in the Appendix~\ref{sec:cora_details}.

\subsection{CORA Generation}
\label{sec:cora_generation}
Unlike UIQ, CORA explicitly introduces \textit{Indirect} queries as an independent interaction type. In realistic retrieval settings, users often employ indirect formulations such as \textit{``Could you help me find \ldots''} or \textit{``I'm trying to locate \ldots''}.

In addition, UIQ relies on fixed opening patterns for each query type, such as \textit{``Can you \ldots''} for Question or \textit{``Locate \ldots''} for Command. Such patterns constrain the query distribution to narrow lexical forms, making it difficult to evaluate robustness to realistic formulation variation. CORA is designed to encourage broader formulation diversity across query types. We additionally verify that generated queries preserve the semantics of the original source captions. Detailed generation procedures, prompts, diversity analyses, and verification results are provided in the Appendix~\ref{sec:cora_details}.

\subsection{Metric}
\label{sec:metrics}

We evaluate CORA reformulations by tracking how the rank of the same target
audio changes when the source query is replaced by a CORA query. For each CORA pair $p \in \mathcal{P}$, let
$\mathrm{rank}_{\mathrm{src}}(p)$ and
$\mathrm{rank}_{\mathrm{cora}}(p)$ denote the 1-indexed rank of the
target audio under the source query and the CORA query, respectively. Both ranks are computed by encoding the corresponding query directly with the retriever's text encoder, with no additional preprocessing or query rewriting applied beforehand. We define rank drop as
\begin{equation}
\mathrm{RD}(p)
=
\mathrm{rank}_{\mathrm{cora}}(p) - \mathrm{rank}_{\mathrm{src}}(p).
\end{equation}
Positive values indicate that the target is ranked lower after reformulation,
while negative values indicate that its rank improves. We report the mean over
all evaluated CORA pairs:
\begin{equation}
\mathrm{MeanRD}
=
\frac{1}{|\mathcal{P}|}\sum_{p \in \mathcal{P}} \mathrm{RD}(p).
\end{equation}
While Recall@5 uses a hard top-$K$ threshold that masks severe absolute rank fluctuations (e.g., dropping from rank 6 to 100), MeanRD performs a paired, instance-level analysis.

\subsection{Embedding-Space Diagnostic Metrics}
\label{sec:embedding_diagnostic_metrics}

Let $q_{\mathrm{src}}$ and $q_{\mathrm{cora}}$ denote the text embeddings of the source query and its CORA reformulation, respectively, and let $a_{\mathrm{tgt}}$ denote the ground-truth target audio embedding. We use cosine similarity as the retrieval score.

\paragraph{Query movement.}
We first define a text-space movement baseline:
\begin{equation}
\Delta_{\mathrm{move}}
=
1 - \cos(q_{\mathrm{cora}}, q_{\mathrm{src}}).
\end{equation}
This metric measures how far the CORA query moves from the source query in the text-embedding space.

\paragraph{Target margin.}
To measure target-margin degradation, we compare the target audio with the
hard negative for each query, defined as the highest-scoring non-target audio:
\begin{equation}
\begin{aligned}
h_{\mathrm{src}}
&=
\arg\max_{a_j \neq a_{\mathrm{tgt}}}
\cos(q_{\mathrm{src}}, a_j), \\
h_{\mathrm{cora}}
&=
\arg\max_{a_j \neq a_{\mathrm{tgt}}}
\cos(q_{\mathrm{cora}}, a_j).
\end{aligned}
\end{equation}

We define the target margin for query $q$ and hard negative $h$ as
\begin{equation}
M(q,h)
=
\cos(q, a_{\mathrm{tgt}}) - \cos(q,h).
\end{equation}
A larger margin means that the target receives a higher score than its hard
negative by a larger amount. We then define Target Boundary Margin Degradation
(TBMD) as the reduction in this margin after reformulation:
\begin{equation}
\mathrm{TBMD}
=
M(q_{\mathrm{src}}, h_{\mathrm{src}})
-
M(q_{\mathrm{cora}}, h_{\mathrm{cora}}).
\end{equation}
A larger TBMD indicates that the CORA query reduces the target's score
advantage over its hard negative.

This margin degradation can be decomposed into two terms:
\begin{equation}
\begin{aligned}
\mathrm{TBMD}
&=
\underbrace{
\cos(q_{\mathrm{src}}, a_{\mathrm{tgt}})
-
\cos(q_{\mathrm{cora}}, a_{\mathrm{tgt}})
}_{\mathrm{TAL}} \\
&\quad+
\underbrace{
\cos(q_{\mathrm{cora}}, h_{\mathrm{cora}})
-
\cos(q_{\mathrm{src}}, h_{\mathrm{src}})
}_{\mathrm{HNP}} .
\end{aligned}
\end{equation}
The first term, Target Alignment Loss (TAL), measures how much similarity to
the target audio decreases after reformulation. The second term, Hard Negative
Pull (HNP), measures how much the hard negative becomes more competitive.
Thus, TBMD captures both sides of retrieval degradation: reduced alignment
with the target and increased competition from a hard negative.
We refer to this formulation as dynamic-HN TBMD. Appendix~\ref{app:hn_sensitivity} verifies that it gives a stronger association with RD than two fixed hard-negative alternatives.

\subsection{Directional Projection Diagnostics}
\label{sec:projection_diagnostics}
We additionally analyze the direction of the source to CORA
query shift in the joint embedding space. We compare its
magnitude $\|d\|_2$ with projections onto the
target-audio direction ($p_{\mathrm{audio}}$) and the
source anchored target boundary direction
($p_{\mathrm{boundary}}$). These are diagnostic associations
with RD, full feature definitions and hard negative selection
are given in Appendices~\ref{sec:proj_math}
and~\ref{sec:proj_hn}.

\section{Experiment}
\label{experiment}

\subsection{Experimental Setup}
\label{sec:experimental_setup}

We evaluate retrieval robustness under the CORA protocol by comparing each source query with five semantically aligned reformulations: Command, Question, Indirect, Key phrase, and Statement. We evaluate four datasets: AudioCaps, Clotho, MACS, and MECAT. For the main diagnostic analysis, retrieval is performed within
each dataset candidate pool. Figure~\ref{fig:overall_query_form_rd}
and Tables~\ref{tab:dataset_form_rankdrop}
and~\ref{tab:dataset_model_rankdrop} use a single candidate pool
containing audio from all four datasets. These results are
reported as a sensitivity analysis.

The main experiments use four CLAP models without additional fine-tuning: LAION-CLAP, M2D-CLAP, MGA-CLAP, and MS-CLAP. For these experiments, we use only
the test splits (3,367 items). Excluding 62 items with an incomplete
set of query forms leaves 3,305 items and yields 66,100 paired
evaluations across five forms and four CLAP retrievers.
Table~\ref{tab:dataset_accounting} provides the dataset-level counts
and filtering details. We additionally apply the same paired protocol
to OEA and RobustCLAP in Section~\ref{sec:external_diagnostics}.
Full dataset and model details are provided in
Appendix~\ref{sec:setup_details}.

\begin{table*}[t]
\centering
\small
\begin{tabular}{lrrrrrr}
\toprule
Dataset & Source captions & Validation & Test & Excluded
        & Retained & Evaluations \\
\midrule
AudioCaps & 1,469 & 495   & 974   & 11 & 963 & 19,260 \\
Clotho    & 2,090 & 1,045 & 1,045 & 51 & 994 & 19,880 \\
MACS      & 1,000 & 500   & 500   & 0  & 500 & 10,000 \\
MECAT     & 1,008 & 160   & 848   & 0  & 848 & 16,960 \\
\midrule
Total     & 5,567 & 2,200 & 3,367 & 62 & 3,305 & 66,100 \\
\bottomrule
\end{tabular}
\caption{Dataset counts and filtering for the main CORA evaluation.
Source captions include the validation and test splits, whereas
retrieval uses only test items. The excluded items have an empty
indirect query (11 AudioCaps items) or an empty key-phrase query
(51 Clotho items). Each retained item contributes five query forms
evaluated with four CLAP retrievers.}
\label{tab:dataset_accounting}
\end{table*}

\subsection{Query-Form Sensitivity Depends on Model-Dataset Regime}
\label{sec:query_form_sensitivity}

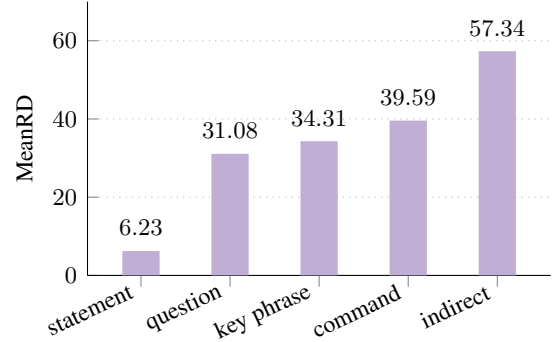
\begin{figure}[t]
\centering
\definecolor{figpurple}{HTML}{C1AED5} 
\definecolor{figorange}{HTML}{FCB563} 

\begin{tikzpicture}
\begin{axis}[
    ybar,
    width=\columnwidth,
    height=5.2cm,
    bar width=14pt,
    ymin=0,
    ymax=70,
    ylabel={MeanRD},
    symbolic x coords={statement, question, key phrase, command, indirect},
    xtick=data,
    x tick label style={rotate=25, anchor=east, font=\small},
    y tick label style={font=\small},
    label style={font=\small},
    axis lines*=left,         
    ymajorgrids=true,         
    grid style={dotted, gray!60}, 
    nodes near coords,
    nodes near coords style={
        font=\footnotesize,
        /pgf/number format/fixed,
        /pgf/number format/precision=2,
        text=black
    },
    every node near coord/.append style={yshift=2pt},
    enlarge x limits=0.15,
]

\addplot[fill=figpurple, draw=none] coordinates {
    (statement, 6.23)
    (question, 31.08)
    (key phrase, 34.31)
    (command, 39.59)
    (indirect, 57.34)
};
\end{axis}
\end{tikzpicture}
\caption{
Overall MeanRD by query form, averaged across all datasets and models. Statement queries (most similar to original captions) are the most stable, while indirect queries cause the largest rank drop.
}
\label{fig:overall_query_form_rd}
\end{figure}
\paragraph{Overall form axis.}
\label{sec:cora_sensitivity_axes}
Figure~\ref{fig:overall_query_form_rd} reports MeanRD by query form, averaged across datasets and CLAP models. Statement queries show the smallest MeanRD on average, likely because their structure is closest to standard caption-style descriptions. In contrast, indirect queries produce the largest rank drop on average, suggesting that this query form is more difficult under the evaluated retrieval settings.

\paragraph{Dataset--form axis.}
Table~\ref{tab:dataset_form_rankdrop} shows that the global hierarchy does not hold uniformly across datasets. For instance, key phrase queries are highly harmful on Clotho, with a MeanRD of 114.92, but actually improve the target rank on MECAT, yielding a MeanRD of $-39.55$. Similarly, statement queries are stable on AudioCaps and Clotho, but cause severe rank drops on MACS. These reversals clearly indicate that CORA sensitivity is not determined by surface form alone; a query form that is safe in one dataset can be harmful in another.

\begin{table}[t]
\centering
\small
\begin{tabular}{llr}
\toprule
\textbf{Dataset} & \textbf{Query form} & \textbf{MeanRD} \\
\midrule
\multirow{5}{*}{AudioCaps}
 & Command    &  -8.13 \\
 & Indirect   &  11.63 \\
 & Key phrase &  23.58 \\
 & Question   & -13.83 \\
 & Statement  & -21.29 \\
\midrule
\multirow{5}{*}{Clotho}
 & Command    &  -1.07 \\
 & Indirect   &  24.19 \\
 & Key phrase & 114.92 \\
 & Question   &   7.01 \\
 & Statement  &  -6.49 \\
\midrule
\multirow{5}{*}{MACS}
 & Command    &  14.66 \\
 & Indirect   &  89.99 \\
 & Key phrase &  20.02 \\
 & Question   &  46.75 \\
 & Statement  &  77.03 \\
\midrule
\multirow{5}{*}{MECAT}
 & Command    & 156.14 \\
 & Indirect   & 128.88 \\
 & Key phrase & -39.55 \\
 & Question   & 101.05 \\
 & Statement  &  10.63 \\
\bottomrule
\end{tabular}
\caption{
Dataset-by-form MeanRD. A query form that causes a large rank drop in one dataset can be neutral or beneficial in another, showing strong dependency on the dataset regime.
}
\label{tab:dataset_form_rankdrop}
\end{table}

\begin{table}[t]
\centering
\small
\begin{tabular}{llr}
\toprule
\textbf{Dataset} & \textbf{Model} & \textbf{MeanRD} \\
\midrule
\multirow{4}{*}{AudioCaps}
 & LAION   &   49.63 \\
 & M2D     &   22.79 \\
 & MGA     &   31.10 \\
 & MS-CLAP & -109.96 \\
\midrule
\multirow{4}{*}{Clotho}
 & LAION   & 31.92 \\
 & M2D     & 17.00 \\
 & MGA     & 16.37 \\
 & MS-CLAP & 45.57 \\
\midrule
\multirow{4}{*}{MACS}
 & LAION   & 214.79 \\
 & M2D     & -80.54 \\
 & MGA     &  -8.15 \\
 & MS-CLAP &  72.67 \\
\midrule
\multirow{4}{*}{MECAT}
 & LAION   & 187.70 \\
 & M2D     &  11.26 \\
 & MGA     &  68.55 \\
 & MS-CLAP &  18.21 \\
\bottomrule
\end{tabular}
\caption{
Dataset-by-model MeanRD under CORA reformulations. The direction and severity
of rank drops vary across CLAP models within the same dataset, indicating that
CORA sensitivity depends on the retriever as well as the dataset.
}
\label{tab:dataset_model_rankdrop}
\end{table}
\paragraph{Dataset--model axis.}
Table~\ref{tab:dataset_model_rankdrop} next decomposes MeanRD by dataset and
CLAP model. This analysis shows that CORA sensitivity also depends on the
retriever. On AudioCaps, LAION, M2D, and MGA all show positive MeanRD,
indicating a rank drop, whereas MS-CLAP shows a large negative MeanRD
($-109.96$). On MACS, LAION and MS-CLAP experience substantial rank drops, while M2D and
MGA show negative MeanRD. MECAT exhibits a different pattern: all four models exhibit rank drops, with LAION showing the largest MeanRD. A complete retriever-level summary comparing aggregate recall and paired RD is provided in Appendix~\ref{sec:retriever_eq}.

\medskip
\noindent\textbf{Summary.} 
Overall, these results show that CORA-induced rank drops cannot be reduced to a single
difficult query form or a single weak model. The direction and severity of
rank change depend heavily on the interaction between the dataset and the retriever. 
The complete dataset--model--query-form retrieval breakdown is provided in Appendix~\ref{app:full_3way_retrieval}.
We further support this interactional pattern with grouped cross-validated prediction in
Appendix~\ref{sec:query_interaction_prediction}.

On the 2,805 items shared with UIQ, we also compare CORA,
CORA without diversification, and released UIQ using identical
source caption anchors, target audios, dataset candidate
pools, and four CLAP retrievers. Their MeanRD values are
+5.28, +4.17, and $-5.78$, respectively. This is a descriptive
comparison of query artifacts with different generation
procedures and form inventories
(Appendix~\ref{app:same_pool_query_sets}).

\subsection{TBMD Characterizes Rank Degradation Better than Query Movement}
\label{sec:embedding_diagnostics}
Section~\ref{sec:query_form_sensitivity} showed that the same query form can be harmful in one model-dataset regime and neutral or beneficial in another.
We therefore ask a more specific diagnostic question: when the target audio drops in rank after CORA reformulation, what kind of embedding-space change explains that degradation?
We compare two accounts.
The \textit{query-movement account} predicts that RD should mainly increase when the reformulated query moves far from the source query in text space.
The \textit{target-boundary account} predicts that RD should increase when the reformulated query weakens the target audio's score advantage over its strongest non-target competitor.
Based on these accounts, we design three diagnostic signals---query movement, Target Alignment Loss (TAL), and Target Boundary Margin Degradation (TBMD)---and test which signal better characterizes continuous target-rank degradation.

\subsubsection{Main Diagnostic Analysis}
\label{sec:tbmd_main_analysis}
\label{sec:main_diagnostic}

We evaluate the above hypothesis over 66,100 CORA pairs from four datasets and four CLAP models.
For each pair, we run retrieval twice over the same candidate pool: once with the source query and once with the CORA query.
This provides the source rank and the CORA rank of the same target audio, allowing us to compute the RD.
We then compute the three diagnostic signals: $\Delta_{\mathrm{move}}$, TAL, and TBMD from the corresponding text and audio embeddings.

\begin{figure}[t]
  \centering
  \includegraphics[width=\columnwidth]{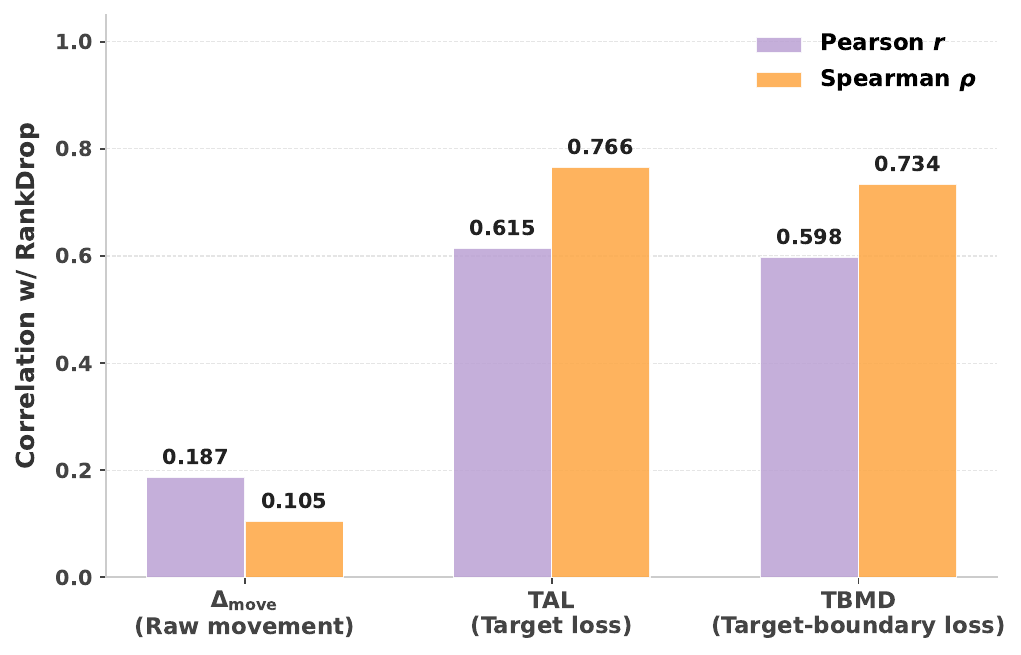}
  \caption{
  Correlation between diagnostic signals and RD over 66,100 CORA pairs using the pooled cross-dataset candidate set.
  Target-specific and boundary-specific signals, TAL and TBMD, show substantially stronger association with RD than query movement $\Delta_{\mathrm{move}}$.
  }
  \label{fig:diagnostic_corr}
\end{figure}

Across the 66,100 CORA pairs evaluated with per-dataset candidate
pools, the overall MeanRD is 3.881. As a sensitivity analysis,
we also evaluate the same pairs with a pooled cross-dataset
candidate set. In this setting, 32,992 pairs (49.91\%) show rank
worsening and MeanRD is 33.711. The pooled-setting correlations are summarized in Figure~\ref{fig:diagnostic_corr}.

\paragraph{Overall diagnostic result.}
In the primary per-dataset analysis, raw query movement ($\Delta_{\mathrm{move}}$) has only a weak association with RD ($r=0.084$). This is a crucial finding: simply reformulating a query so that it moves far away from the original caption in text space does not automatically mean the target audio will be lost.

By contrast, the target-boundary signals show much stronger associations with rank degradation. Target Alignment Loss (TAL) and Target Boundary Margin Degradation (TBMD) achieve Pearson correlations of 0.508 and 0.615, respectively. This supports the target-boundary account: rank degradation is associated with a loss of target alignment or a narrowing of the target's margin against competitors, rather than with how far the text moves on the surface. Because TBMD is computed from the same similarity scores that determine rank, this association is diagnostic rather than causal; Appendix~\ref{app:tbmd_control} reports a control analysis showing that TBMD retains explanatory power over TAL after accounting for dataset, model, query form, $\Delta_{\mathrm{move}}$, source rank, and candidate pool size. Appendix~\ref{app:anchor_content_robustness} further checks anchor and content controls, and Appendix~\ref{sec:token_saliency_appendix} supports this interpretation with a token-level audit.

On the matched three-dataset subset, the released UIQ queries show the same ordering: Pearson's $r$ with RankDrop is $-0.075$ for $\Delta_{\mathrm{move}}$, $0.508$ for TAL, and $0.538$ for dynamic-HN TBMD (Appendix~\ref{app:same_pool_query_sets}). This is a descriptive comparison because UIQ differs in generation procedure, grounding, and query-type inventory.

\paragraph{Dataset-level analysis.}
In the pooled cross-dataset sensitivity analysis, we further examine whether this diagnostic pattern holds across datasets.
Here, Pearson's $r$ measures the linear association between each diagnostic signal and RD:
values near zero indicate little association, whereas larger positive values indicate that the signal increases with RD.
Across AudioCaps, Clotho, MACS, and MECAT, the association for $\Delta_{\mathrm{move}}$ tends to be lower or more variable, whereas TAL and TBMD generally exhibit higher positive associations with RD.
For example, $\Delta_{\mathrm{move}}$ shows little correlation with RD on AudioCaps ($r=-0.016$), while TAL and TBMD reach $r=0.531$ and $r=0.547$.
On MACS, $\Delta_{\mathrm{move}}$ reaches $r=0.361$, whereas TAL and TBMD reach $r=0.737$ and $r=0.716$.
These dataset-level results indicate that the diagnostic pattern is not isolated to a single dataset:
RD tends to align more closely with TAL and TBMD than with $\Delta_{\mathrm{move}}$.
The full dataset-level breakdown is provided in Appendix~\ref{app:dataset_level_diagnostic}.

\subsubsection{TBMD-Guided View Intervention}
\label{sec:tbmd_intervention}

The previous analysis shows that TBMD is consistently associated with RD.
We further conduct a controlled view-selection check: for each source caption,
we select one of the five CORA reformulations while keeping the target audio,
retrieval model, and candidate pool fixed. If TBMD captures the degradation
pattern, \textbf{Min-TBMD oracle} should yield lower MeanRD than the
text-space-closest policy.

\noindent\textbf{Selection policies.}
\begin{itemize}[leftmargin=*, nosep]
    \item \textbf{All CORA forms}: average over all five reformulations.
    \item \textbf{Min-$\boldsymbol{\Delta}_{\mathrm{move}}$ view}: choose the text-space-closest reformulation.
    \item \textbf{Min-TAL oracle}: choose the reformulation with the smallest TAL.
    \item \textbf{Min-TBMD oracle}: choose the reformulation with the smallest TBMD.
    \item \textbf{Best-rank oracle}: choose the reformulation with the best target rank.
\end{itemize}

The \textbf{Min-TAL oracle}, \textbf{Min-TBMD oracle}, and \textbf{Best-rank oracle} are target-aware diagnostic oracles and are not intended as inference-time methods.

\begin{table}[t]
\centering
\small
\setlength{\tabcolsep}{6pt}
\begin{tabular}{lr}
\toprule
\textbf{Selection policy}
& \textbf{MeanRD} \\
\midrule
\textbf{Average over views} &   3.881 \\
\textbf{Min-$\boldsymbol{\Delta}_{\mathrm{move}}$ view} &   1.875 \\
\textbf{Min-TAL oracle} & -18.568 \\
\textbf{Min-TBMD oracle} & -19.467 \\
\textbf{Best-rank oracle} & -24.560 \\
\bottomrule
\end{tabular}
\caption{
Controlled CORA-view intervention. Lower MeanRD indicates better target rank
preservation; negative values indicate that the selected CORA view improves the
target rank relative to the source query. Values in this table are used to
compare selection policies under the same controlled setting.
}
\label{tab:tbmd_intervention}
\end{table}

Table~\ref{tab:tbmd_intervention} shows that selecting the
\textbf{Min-$\boldsymbol{\Delta}_{\mathrm{move}}$ view} only slightly reduces
MeanRD, from 3.881 to 1.875, compared with averaging over CORA views.
In contrast, \textbf{Min-TBMD oracle} reduces MeanRD to $-19.467$,
slightly below \textbf{Min-TAL oracle} ($-18.568$) and closer to the
\textbf{Best-rank oracle} upper bound ($-24.560$).
This is consistent with the diagnostic role of TBMD: compared with TAL alone,
TBMD also accounts for whether a hard negative becomes more competitive after
reformulation.

\begin{figure}[t]
    \centering
    \includegraphics[width=\linewidth]{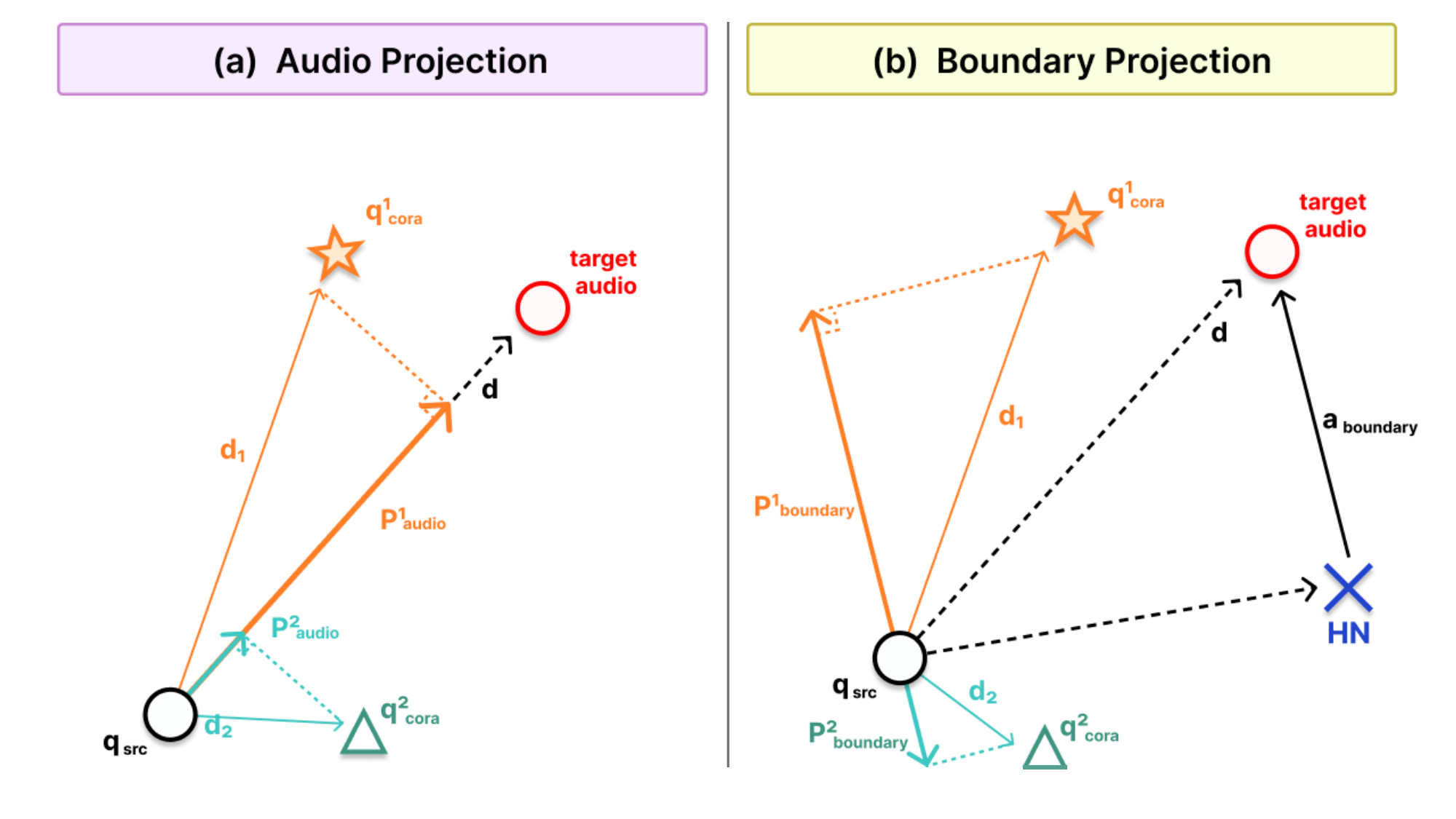}
    \caption{
    Illustration of (a) the audio-projection and (b) the boundary-projection perspectives for query shift.
    }
    \label{fig:projection_analysis}
\end{figure}

\subsection{Projection Analysis Validates the Directional Boundary Hypothesis}

The previous section analyzed target-alignment change and local-margin degradation using scalar metrics such as TAL and TBMD. However, these scalar quantities do not fully reveal whether the retrieval degradation is caused by the direction of the embedding shift. Therefore, we further analyze whether the embedding shift from the source caption to the CORA query moves in a direction that preserves or degrades the retrieval boundary between the target and hard negative.

We analyze the relationship between RD and three features: $||d||_2$, $p_{\mathrm{audio}}$, and $p_{\mathrm{boundary}}$. Formal feature definitions and hard negative selection details are provided in the Appendix~\ref{sec:proj_math} and Appendix~\ref{sec:proj_hn}

\begin{itemize}
    \item $\|d\|$: the magnitude of the embedding shift, indicating how far the reformulated query moves in the embedding space.
    \item $p_{\mathrm{audio}}$: the projection of $d$ onto the target-audio direction, measuring alignment between the query movement and the target audio embedding.
    \item $p_{\mathrm{boundary}}$: the projection of $d$ onto the local target boundary direction, measuring whether the query movement preserves retrieval boundary against nearby hard negatives.
\end{itemize}

Figure~\ref{fig:projection_analysis} illustrates the
core intuition behind the projection analysis.
Figure~\ref{fig:projection_analysis}(a) presents
the audio-projection perspective. Although
$q_{\mathrm{cora}}^{1}$ moves farther away from the
source caption embedding than
$q_{\mathrm{cora}}^{2}$, its embedding shift direction
$d_1$ remains better aligned with the target-audio
direction, resulting in less retrieval collapse.

Figure~\ref{fig:projection_analysis}(b) presents
the boundary-projection perspective. Even with
larger displacement, $q_{\mathrm{cora}}^{1}$ remains
better aligned with the target
boundary direction, thereby preserving local
retrieval separation against nearby hard negatives.
Meanwhile, $q_{\mathrm{cora}}^{2}$ weakens the local
retrieval boundary despite its smaller $||d||_2$

To evaluate the explanatory power of each feature set, we perform grouped cross-validated Ridge regression for RD prediction. Detailed regression settings are provided in the Appendix~\ref{sec:proj_regression} We then compute Pearson and Spearman correlations between predicted and ground-truth RD values.

\begin{table}[t]
\centering
\small
\begin{tabular}{lcc}
\toprule
\textbf{Feature set} & \textbf{Pearson $r$} & \textbf{Spearman $\rho$} \\
\midrule
fixed effects only & 0.174 & 0.178 \\
+ $\|d\|$ & 0.206 & 0.196 \\
+ $p_{\mathrm{audio}}$ & 0.500 & 0.594 \\
+ $p_{\mathrm{boundary}}$ & 0.422 & 0.529 \\
\bottomrule
\end{tabular}
\caption{Correlation between projection-based feature sets and RD prediction.}
\label{tab:projection_features}
\end{table}

As shown in Table~\ref{tab:projection_features}, all regression settings additionally include fixed effects for dataset, retrieval model, and query form. Adding the $\|d\|$ achieves only a weak association with RD ($\rho = 0.196$).  In contrast, the projection-based features show substantially stronger correlations: $p_{\mathrm{audio}}$ reaches $\rho = 0.594$, while $p_{\mathrm{boundary}}$ reaches $\rho = 0.529$.

These results support the TBMD interpretation from a directional perspective: retrieval degradation occurs not when the query moves far from the source, but when the embedding shift fails to preserve target alignment and local boundary separation from hard negatives.

\subsection{TBMD Characterizes Residual Rank Degradation in Query-Robust Retrievers}
\label{sec:external_diagnostics}

We further test whether the target-boundary pattern observed above also holds for retrievers designed or evaluated for query-form robustness.
We evaluate three external retrievers: OEA-Qwen3B-Cl, OEA-Qwen3B-AC, and RobustCLAP.
OEA represents user-intent-oriented audio-text retrieval, and RobustCLAP represents robustness-oriented retrieval trained with paraphrastic views of the same audio scene \citep{oea2025,selvakumar2025robustclap}.
We ask whether the remaining CORA rank degradation in these models is still better characterized by TBMD than by query movement.
To investigate this, we apply the CORA paired evaluation protocol to each external retriever, computing RD and diagnostic correlations independently.

\begin{table*}[t]
\centering
\small
\setlength{\tabcolsep}{5pt}
\resizebox{0.86\textwidth}{!}{%
\begin{tabular}{lrrrr}
\toprule
\textbf{Model}
& \textbf{CORA R@5}
& \textbf{MeanRD}
& $\boldsymbol{r(\Delta_{\mathrm{move}},\mathrm{RD})}$
& $\boldsymbol{r(\mathrm{TBMD},\mathrm{RD})}$ \\
\midrule
OEA-Qwen3B-Cl & 43.34 & 0.603 & 0.084 & 0.472 \\
OEA-Qwen3B-AC & 41.17 & 0.479 & 0.046 & 0.478 \\
RobustCLAP    & 35.47 & 9.408 & 0.174 & 0.508 \\
\bottomrule
\end{tabular}
}
\caption{
External retriever diagnostics under the CORA protocol.
CORA R@5 is reported as a percentage.
MeanRD measures average target-rank degradation after reformulation.
Correlations are computed separately for each retriever.
}
\label{tab:external_diagnostics_tbmd}
\end{table*}

Table~\ref{tab:external_diagnostics_tbmd} shows two patterns.
First, OEA variants substantially reduce CORA degradation, with MeanRD close to zero.
RobustCLAP shows larger MeanRD in this evaluation, despite being designed for linguistic-variation robustness.
Second, the residual rank degradation in all three retrievers follows the same diagnostic ordering as in the CLAP models.
Query movement is only weakly associated with RD, with correlations from 0.046 to 0.174.
By contrast, TBMD is more strongly associated with RD across all three retrievers, with correlations from 0.472 to 0.508.

This result indicates that models designed for intent-oriented or linguistic-variation-robust retrieval can reduce the severity of CORA degradation, but the remaining errors are still explained by the same target-boundary mechanism.
In other words, even after robustness-oriented training or evaluation, residual CORA RD is better characterized by degradation of the target's margin against its strongest non-target competitor than by the amount of query movement alone.

\section{Discussion}
\label{sec:discussion}

CORA should be interpreted as a paired diagnostic protocol. Recall@$k$ measures
aggregate retrieval quality, while RD checks whether the same target audio
is preserved after a CORA reformulation. This paired view is important because
our results show that query form sensitivity changes across the model and
dataset regime, rather than following a fixed ordering of surface forms.

The embedding diagnostics suggest that residual RD is not well explained
by raw text space movement alone. TAL and TBMD provide more informative signals
because they measure whether the reformulated query weakens target alignment and
the target audio's score advantage over its Hard Negative. These results do not
show that TAL or TBMD should be used as inference time objectives. Rather, they
indicate which retrieval changes are associated with CORA failures under fixed
targets and candidate pools.


This diagnostic perspective clarifies the limits of simple inference time corrections. As shown in Appendix~\ref{sec:fusion_check}, interventions like MultiCLAP score fusion can improve aggregate R@5 as a supporting baseline, but they do not resolve the underlying query form brittleness. Appendix~\ref{sec:preprocessing} finds a similar result for query preprocessing. Compared with no preprocessing, general-purpose or CORA-specific rewriting lowers MeanRD, but the TAL/TBMD over movement ordering holds under all three conditions.
True robustness requires explicitly addressing intent variation during the training stage, as demonstrated by OEA's reduced RD. Therefore, future robustness methods must be evaluated not just on aggregate performance, but on their ability to preserve a target's margin over competitive non target audio under expressive reformulation.

\section{Conclusion}

We introduced CORA, a caption-anchored diagnostic protocol for evaluating
whether text-to-audio retrievers preserve the same target audio under
intent-preserving query reformulations. Across four datasets and multiple
retrievers, CORA shows that query-form sensitivity is conditional on the
model-dataset regime rather than being an intrinsic property of a particular
surface form.

Our main diagnostic finding is that residual RD is more closely associated
with Target Alignment Loss and Target Boundary Margin Degradation than with raw
text-space movement. The same pattern also appears in external robustness-oriented
retrievers, including OEA and RobustCLAP. These results suggest that robust T2A
retrieval requires more than invariance to query wording: it should preserve the
target audio's score advantage over competitive non-target audio under
reformulation. CORA provides a compact testbed for analyzing this target-boundary
behavior and for developing future boundary-aware robustness methods.

\section*{Limitations}

This work has three main limitations. First, CORA uses LLM-generated
reformulations rather than real user search logs; although we verify semantic
alignment and query type validity, generated queries may still contain artifacts
and underrepresent multilingual queries. Second, CORA is a
controlled caption reformulation diagnostic, so it does not cover unconstrained
intent, multi-target relevance, or specialized domains such as speech, music,
medical, or industrial audio. Third, our findings are based on CLAP models and
selected external retrievers, and TAL/TBMD remain diagnostic rather than causal
signals whose behavior may change with cross-encoder, autoregressive, or
larger-corpus retrieval settings and with different Hard Negative pools.

\section*{Acknowledgments}
This work was supported by the Institute of Information \&
Communications Technology Planning \& Evaluation(IITP) grant
funded by the Korea government(MSIT) (No. RS-2025-25441313,
Professional AI Talent Development Program for Multimodal AI
Agents). This research was also supported by the MSIT, Korea,
under the National Program for Excellence in AI University
(Sogang University) supervised by the IITP in 2026.
(Grant No. RS-2026-0-00036)

\bibliography{custom}

\clearpage
\appendix
\setlength{\tabcolsep}{4pt}
\renewcommand{\arraystretch}{1.05}

\section{CORA Details}
\label{sec:cora_details}
\subsection{CORA taxonomy}
\label{sec:cora_taxnanomy}

\begin{itemize}

\item \textbf{Command Query}: Imperative queries that directly request detection or retrieval of a target sound (e.g., \textit{``Detect a dog barking while vehicles are approaching and then driving past on a nearby road.''})

\item \textbf{Question Query}: Natural language questions asking whether a specific sound event exists (e.g., \textit{``Are there sounds of a dog barking while vehicles approach and then drive past on a nearby road?''})

\item \textbf{Indirect Query}: Indirect or polite request-style queries commonly used in conversational interactions (e.g., \textit{``At your convenience, could you help locate the sound where a dog is barking while vehicles are approaching and then driving past on a nearby road?''})

\item \textbf{Key phrase Query}: Keyword-centric queries with minimal sentence structure (e.g., \textit{``dog barking with approaching and passing vehicles on nearby road''})

\item \textbf{Statement Query}: Declarative queries that objectively describe sound events without explicit requests or emotional expressions (e.g., \textit{``A dog is barking while vehicles approach and then drive past on a nearby road.''})

\end{itemize}

\subsection{CORA Generation}

CORA was generated using GPT-4.1 with temperature $=1.0$, top-p=0.9, and maximum generation length of 2048 tokens. For each query type, we designed different personas and generation constraints to reflect the diversity of real-world user search expressions. We construct CORA from the test and validation splits of four audio-text retrieval datasets: Clotho (1,045 test and 1,045 validation samples), AudioCaps (974 test and 495 validation samples), MECAT (848 test and 160 validation samples), and MACS (500 test and 500 validation samples), resulting in a total of 5,567 source captions.

The 5,567 source captions include 3,367 test and 2,200
validation captions. Retrieval evaluation uses only the test
split. The main retrieval analysis uses only test items with all five non-empty query forms. This removes 11 AudioCaps items with an empty indirect query and 51 Clotho items with an empty key-phrase query. The resulting 3,305
items form the complete evaluation set.

During the generation of indirect, command, and question queries, we observed that query openings often converged to repetitive patterns. To alleviate this issue, CORA introduces a training-free diversification prompting strategy.

The motivation for this approach is based on the observation that LLM generation can be viewed as probabilistic sampling conditioned on prompts. Prior work~\cite{zhang2025verbalized} suggests that prompts can induce different response distributions. We hypothesized that frequently repeated opening expressions correspond to high-probability regions of the generation distribution. Based on this intuition, we first force the model to generate multiple plausible opening expressions and then explicitly prohibit the final query from using any of the generated candidates. This process encourages the model to avoid repeatedly generated prefixes and increases formulation diversity.

Specifically, before generating the final query, the model is first instructed to generate multiple plausible opening expressions for the target query type. The model is then required to select a final opening expression that does not appear in the generated candidate list. By forcing the final query to begin with a non-listed expression, the generation process reduces prefix collapse and increases formulation diversity across query types.

The prompts used for each query type are provided in Appendix ~\ref{tab:querydefinition}.

\subsection{CORA Verification}
\subsubsection{Query Quality Verification}
\label{sec:verification}
\paragraph{LLM-as-a-Judge and Disjoint Human Verification.}
We validate CORA queries along two axes: \textit{semantic alignment} (whether the original caption's meaning is preserved after reformulation) and \textit{query-type validity} (whether each query correctly reflects its intended type).
We evaluate 100 samples from each dataset (AudioCaps, Clotho, MACS, and MECAT), resulting in a total of 500 evaluated samples, using two LLM judges---Gemini 2.5 Flash and Qwen2.5-72B-Instruct---both of which exhibit human-like agreement patterns rather than super-consistent behavior~\citep{han2025judgesverdict}, making them well-suited for subjective judgment tasks.
Each judge is run three times, showing stable overall performance with variation within 0.3 percentage points across runs.
Human verification is additionally conducted by five undergraduate annotators with high English proficiency.

As shown in Table~\ref{tab:verification_summary}, all query types achieve high scores on both axes across all datasets, with most types exceeding 90\% . Human verification yields 98.3\% on semantic alignment and 97.7\% on query-type validity, consistent with the automated results.
These findings confirm that CORA queries are semantically faithful and linguistically well-formed with respect to their target types.

For human verification, we recruit 5 undergraduate students with high English proficiency as annotators. All annotators participated as volunteers without compensation and were informed of the research purpose before participation. Each annotator is assigned to a disjoint subset of samples to independently assess semantic alignment and query type validity.

\begin{table*}[t]
\centering
\footnotesize
\setlength{\tabcolsep}{3pt}
\renewcommand{\arraystretch}{0.92}
\begin{adjustbox}{max width=.78\textwidth}
\begin{tabular}{lcccccc}
\toprule
& \multicolumn{2}{c}{\textbf{Human}} & \multicolumn{2}{c}{\textbf{Gemini 2.5 Flash}} & \multicolumn{2}{c}{\textbf{Qwen2.5-72B}} \\
\cmidrule(lr){2-3} \cmidrule(lr){4-5} \cmidrule(lr){6-7}
\textbf{Query Type} & \textbf{T1} & \textbf{T2} & \textbf{T1} & \textbf{T2} & \textbf{T1} & \textbf{T2} \\
\midrule
Key Phrase & 97.6 & 97.8  & 89.7 & 100.0 & 100.0 & 100.0  \\
Statement  & 97.0 & 100.0 & 100.0 & 100.0 & 100.0 & 100.0  \\
Question   & 99.4 & 94.9  & 92.0 & 97.7 & 100.0 & 90.3 \\
Command    & 98.2 & 100.0 & 93.3 & 95.0 & 95.3 & 96.0  \\
Indirect   & 99.4 & 95.6  & 99.0 & 100.0 & 100.0 & 100.0 \\
\midrule
Overall    & 98.3 & 97.7  & 94.8 & 98.5 & 99.1 & 97.3 \\
\bottomrule
\end{tabular}
\end{adjustbox}

\caption{
  Verification accuracy (\%) for semantic alignment (Task~1) and query-type validity (Task~2)
  across five CORA query types.
}
\label{tab:verification_summary}
\end{table*}

\paragraph{Overlapping Validation and Inter-Annotator Reliability.}

To address potential individual annotator bias and evaluate inter-annotator agreement (Cohen's $\kappa$), we conducted an additional overlapping human validation on 100 CORA reformulation pairs.
All five annotators independently assessed the exact same 100 pairs across both Task 1 (Semantic Alignment) and Task 2 (Query-Type Validity) using the identical rubric provided to the LLM judges.


As summarized in Table~\ref{tab:human_overlap_results}, annotators demonstrated strong inter-annotator reliability, achieving a mean pairwise Cohen's $\kappa$ of 0.970 (range: 0.925--1.000) with 98.8\% accuracy for Task 2, and 98.2\% raw agreement for Task 1.

\begin{table}[h]
\centering
\small
\begin{tabular}{lc}
\toprule
\textbf{Metric / Evaluation Axis} & \textbf{Value} \\
\midrule
Task 1 Raw Agreement (Semantic Alignment) & 98.2\% \\
Task 2 Accuracy (Query-Type Validity)     & 98.8\% \\
Mean Pairwise Cohen's $\kappa$            & 0.970  \\
Pairwise Cohen's $\kappa$ Range           & 0.925--1.000 \\
\bottomrule
\end{tabular}

\caption{Summary of additional overlapping human validation across 5 annotators on the same 100 pairs.}

\label{tab:human_overlap_results}
\end{table}

Out of 500 judgments in Task 2, only 6 single-annotator errors occurred. As detailed in Table~\ref{tab:human_error_analysis}, these discrepancies were mostly subtle boundary cases between interrogative/indirect forms and key phrases vs. imperatives.

\begin{table}[h]
\centering
\small
\begin{tabular}{llc}
\toprule
\textbf{Error Type} & \textbf{Confusion Pair} & \textbf{Count} \\
\midrule
Interrogative form      & Question $\leftrightarrow$ Indirect & 3 \\
Phrase vs. sentence     & Key phrase $\leftrightarrow$ Statement & 1 \\
Phrase vs. imperative   & Key phrase $\leftrightarrow$ Command & 1 \\
Declarative vs. indirect & Statement $\leftrightarrow$ Indirect & 1 \\
\midrule
\textbf{Total}          & & \textbf{6} \\
\bottomrule
\end{tabular}
\caption{Breakdown of query-type confusion errors observed in the overlapping human validation.}
\label{tab:human_error_analysis}
\end{table}

\subsubsection{Five-Form Taxonomy Validation}
\label{sec}

To examine whether our findings depend on any particular query form, we conduct a leave-one-form-out analysis. Specifically, we remove each of the five query forms and recompute the correlations between RankDrop and TBMD, TAL, and $\Delta_{\mathrm{move}}$ using the remaining four forms. Across all five leave-one-form-out conditions, both TBMD ($r=0.607$--$0.625$) and TAL ($r=0.494$--$0.526$) remain substantially more strongly correlated with RankDrop than $\Delta_{\mathrm{move}}$ ($r=0.060$--$0.098$). Thus, the TAL/TBMD $\gg$ $\Delta_{\mathrm{move}}$ pattern is preserved regardless of which query form is excluded, indicating that our main diagnostic conclusion is not driven by any single query form.

\subsubsection{Real-World Query Coverage}

To examine whether our five query forms reflect how users express information needs in practice, we additionally analyze WildClaims, a dataset of real conversational user interactions. We focus on its \textit{Information seeking} subset, which contains queries where users ask for specific information or facts and thus provides a practical proxy for user queries. We classify all 2,538 information-seeking queries into our five forms (\textit{Question}, \textit{Command}, \textit{Key Phrase}, \textit{Statement}, and \textit{Indirect}) plus a  \textit{None} category using Gemini 2.5 Flash with deterministic decoding ($T=0$).

Overall, 97.6\% of the real-user queries are covered by our five forms: 55.7\% are Questions, 23.5\% Commands, 9.6\% Key Phrases, 5.2\% Statements, and 3.6\% Indirect queries. Only 2.4\% are classified as \textit{None}. Most of these cases do not exhibit a clear or typical query form, including copied search-result JSON, webpage dumps, or incomplete fragments. Qualitative examples in Table~\ref{tab:real-user-query} further illustrate that the same five query forms occur in real-user queries. The classification prompt is provided in Appendix~\ref{app:wildclaims-prompt}.

\begin{table*}[t]
\centering
\small
\begin{tabular}{p{0.11\textwidth} p{0.40\textwidth} p{0.40\textwidth}}
\toprule
\textbf{Form} & \textbf{Real-user query (WildClaims)} & \textbf{CORA query} \\
\midrule

Key Phrase &
Free resources to get AMHS based messages &
Crowd speech laughter \\

Statement &
The bomb was dropped on Hiroshima at around 8 am. &
Many people are speaking and laughing. \\

Question &
Can you find leet cheatsheet? &
Does the recording feature pigeons cooing, bird wings flapping, gravel shuffling, and wood clacking? \\

Command &
Show me sources behind all of these bullets. &
Retrieve speech and laughter from many people. \\

Indirect &
Can you please provide me with citable sources of journals that discusses the topic of organizational innovation technology adoption and the inconsistencies between research studies. &
Would you be able to help me locate the audio where a train is running on train tracks and a steam engine horn is whistling? \\

\bottomrule
\end{tabular}
\caption{Examples of real-user information-seeking queries from WildClaims and CORA queries exhibiting the same query forms.}
\label{tab:real-user-query}
\end{table*}

These results provide empirical evidence that the five-form taxonomy covers most real-world information-seeking queries. We do not aim to reproduce their natural frequency distribution. Instead, CORA uses balanced reformulations to isolate the effect of query form on retrieval behavior.

\subsection{Additional Diversity Analysis}

\subsubsection{Experimental Setup}

To analyze whether CORA increases query diversity in practice, we compare UIQ, CORA(w/o Prompt), and CORA.

UIQ is the original query benchmark. CORA(w/o Prompt) uses the same taxonomy as CORA but removes the diversification strategy. CORA is the full version proposed in this work.

For fair comparison, we sampled 100 examples each from AudioCaps, Clotho, and MECAT, resulting in approximately 300 queries per query type. MACS was excluded because it is not included in UIQ.

Sentence embeddings were extracted using \texttt{sentence-transformers/all-MiniLM-L6-v2}.

\subsubsection{Metrics}

We measure semantic diversity using a pairwise cosine-distance-based diversity metric. Global diversity is computed as the average cosine distance between all query embeddings:

\[
D_{\mathrm{global}}
=
\mathrm{mean}_{i<j}
\left(
1 - e(q_i)^\top e(q_j)
\right)
\]

where \(i\) and \(j\) index different queries, and \(e(q_i)\) denotes the normalized sentence embedding of query \(q_i\). Higher global diversity indicates that query formulations are more widely distributed in the embedding space.

To additionally measure formulation diversity within the same query type, we compute within-type diversity:

\[
D_{\mathrm{within}}
=
|T|^{-1}
\sum_{t \in T}
\mathrm{mean}_{i<j}
\left(
1 - e(q_i^t)^\top e(q_j^t)
\right)
\]

where \(q_i^t\) and \(q_j^t\) denote two different queries belonging to query type \(t\). Since the diversification prompting strategy is applied only to command, indirect, and question queries, within-type diversity is evaluated exclusively on these query types.
 
\subsubsection{UIQ vs. CORA vs. CORA(w/o Prompt)}

\begin{figure}[t]
  \centering
  \includegraphics[width=\columnwidth]{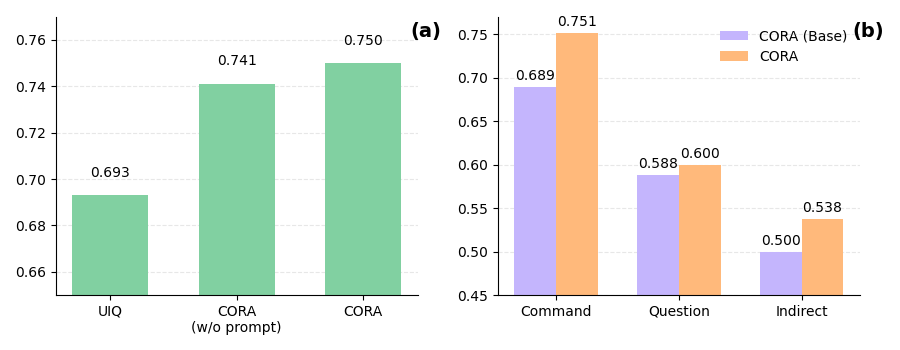}
  \caption{
  Semantic diversity comparison between UIQ, CORA(w/o Prompt), and CORA.
  (a) Global semantic diversity across all query embeddings.
  (b) Within-type semantic diversity for command, question, and indirect queries.
  CORA consistently improves diversity over CORA(w/o Prompt), with the largest gain observed for the command type.
  }
  \label{fig:semantic_diversity}
\end{figure}

Compared to UIQ, CORA(w/o Prompt) achieves substantially higher global semantic diversity (0.693 $\rightarrow$ 0.741; Figure~\ref{fig:semantic_diversity}(a)). This suggests that introducing indirect query types expands the overall embedding distribution by incorporating conversational and request-oriented formulations that occupy regions distinct from existing query styles.

CORA further improves global diversity over CORA(w/o Prompt) (0.741 $\rightarrow$ 0.750; Figure~\ref{fig:semantic_diversity}(a)). Since both variants share the same query taxonomy, the improvement mainly comes from the proposed diversification prompting strategy rather than query-type expansion itself.

Within-type analysis further shows that the command type benefits the most from the diversification strategy (0.689 $\rightarrow$ 0.751; Figure~\ref{fig:semantic_diversity}(b)). In contrast, question queries exhibit relatively smaller gains because yes/no question structures naturally constrain formulation variability. 

\subsection{Same-Pool Retrieval Comparison}
\label{app:same_pool_query_sets}

We compare CORA, CORA without diversification, and released UIQ
on the same 2,805 test items: 963 from AudioCaps, 994 from
Clotho, and 848 from MECAT. MACS is not included in UIQ.
The evaluation uses the same source captions, target audios,
per-dataset candidate pools, and four CLAP retrievers.

\begin{table}[t]
\centering
\small
\begin{tabular}{lrrr}
\toprule
Query set & Forms & Evaluations & MeanRD \\
\midrule
CORA          & 5 & 56,100 & +5.28 \\
CORA w/o div. & 5 & 56,100 & +4.17 \\
Released UIQ  & 4 & 44,880 & $-5.78$ \\
\bottomrule
\end{tabular}
\caption{MeanRD for three query sets evaluated on the same
2,805 items and per-dataset candidate pools.}
\label{tab:same_pool_query_sets}
\end{table}

\begin{table}[t]
\centering
\small
\setlength{\tabcolsep}{3pt}
\resizebox{\columnwidth}{!}{%
\begin{tabular}{lccc}
\toprule
Query set ($N$) & $\Delta_{\mathrm{move}}$ & TAL & Dynamic-HN TBMD \\
\midrule
CORA (56,100) & .092/.071 & .457/.574 & .535/.667 \\
CORA w/o diversification (56,100) & .081/.064 & .443/.578 & .519/.660 \\
Released UIQ (44,880) & $-.075/-.073$ & .508/.669 & .538/.736 \\
\bottomrule
\end{tabular}%
}
\caption{Pearson $r$/Spearman $\rho$ correlations with RankDrop under matched per-dataset candidate pools. $N$ counts item-by-query-form-by-retriever observations across 2,805 items and four retrievers. This is a descriptive comparison because query generation, grounding, and query-type inventories differ.}
\label{tab:same_pool_diagnostic_corr}
\end{table}

The paired MeanRD difference is +1.10 for CORA versus CORA
without diversification (95\% cluster-bootstrap CI [0.75, 1.45])
and +11.06 for CORA versus UIQ ([9.35, 12.82]).
These are descriptive comparisons because the query sets differ
in generation conditions and, for UIQ, query types. The generator
also differs: CORA and CORA without diversification were generated
with GPT-4.1, whereas UIQ was generated with GPT-5.1. All three
query sets preserve the TAL/TBMD-over-$\Delta_{\mathrm{move}}$
ordering, so the pattern is not confined to GPT-4.1-generated
queries, although this is not a controlled generator-stability
experiment.

\section{Experimental Setup Details}
\label{sec:setup_details}
 
\noindent\textbf{Datasets.}
We evaluate retrieval robustness on AudioCaps, Clotho, MACS, and MECAT.
AudioCaps is based on AudioSet, and several clips overlap with the WavCaps AudioSet\_SL subset.
Similarly, Clotho is built on Freesound, which substantially overlaps with the WavCaps Freesound subset.
Such overlap may lead to overestimated retrieval performance.
MECAT is constructed from ACAV100M web-video audio data and does not directly overlap with WavCaps sources.
MACS is based on TAU Urban Acoustic Scenes 2019 and consists of real-world urban scene recordings.
Therefore, MECAT and MACS can be regarded as relatively leakage-free benchmarks.
 
\noindent\textbf{Retrieval Models.}
We evaluate four CLAP-family retrievers without additional fine-tuning:
\begin{itemize}
    \item \textbf{LAION-CLAP}: a representative zero-shot T2A retrieval model based on contrastive audio-language pretraining.
    \item \textbf{M2D-CLAP}: a retrieval model utilizing masked modeling-based representation learning.
    \item \textbf{MGA-CLAP}: a retrieval model designed for multi-granularity audio-text alignment.
    \item \textbf{MS-CLAP}: a CLAP variant trained with a multi-stage training strategy.
\end{itemize}

\section{Retriever-Level CORA Evaluation}
\label{sec:retriever_eq}
Table~\ref{tab:retriever_eq_rankdrop} provides the global marginal retrieval performance and paired rank degradation (MeanRD) across all evaluated retrievers, including both CLAP models and external robustness-oriented models.
 As discussed in Section~\ref{sec:query_form_sensitivity}, analyzing aggregate $\Delta$R@5 alone can be misleading because it obscures instance-level target retention. For example, while MS-CLAP shows a positive aggregate $\Delta$R@5 (+3.07\%), this increase starts from a very low source caption baseline (17.37\%); its MeanRD remains near zero, indicating that the original targets are not necessarily better preserved. Similarly, external models like RobustCLAP show severe paired rank drops (MeanRD 9.408) despite their robustness-oriented training, which aggregate metrics might partially mask. This comprehensive retriever-level summary supports our methodological choice to rely on paired, continuous RD metrics rather than aggregate top-$k$ recall changes.

\begin{table*}[t]
\centering
\small
\setlength{\tabcolsep}{8pt}
\begin{tabular}{llrrrr}
\toprule
\textbf{Retriever} & \textbf{Type} & \textbf{Source R@5} & \textbf{CORA R@5} & $\boldsymbol{\Delta}$\textbf{R@5} & \textbf{MeanRD} \\
\midrule
LAION & single & 29.68 & 22.17 & -7.51 & 104.71 \\
MS-CLAP & single & 17.37 & 20.44 &  3.07 & -2.67 \\
M2D & single & 34.89 & 29.57 & -5.31 &  2.46 \\
MGA & single & 34.52 & 31.69 & -2.83 & 30.34 \\
MultiCLAP-ScoreAvg-Z & fusion & 38.67 & 35.75 & -2.92 & 24.29 \\
MultiCLAP-ScoreAvg-MinMax & fusion & 38.28 & 35.50 & -2.77 & 25.04 \\
OEA-Qwen3B-Cl & external & 44.01 & 43.34 & -0.67 & 0.603 \\
OEA-Qwen3B-AC & external & 42.04 & 41.17 & -0.87 & 0.479 \\
RobustCLAP & external & 47.24 & 35.47 & -11.77 & 9.408 \\
\bottomrule
\end{tabular}
\caption{
Comprehensive retriever-level performance under the CORA protocol.
R@5 and $\Delta$R@5 values are reported as percentages.
MeanRD is computed over all CORA pairs.
This table illustrates the limitation of aggregate $\Delta$R@5 and motivates the use of paired MeanRD to accurately capture instance-level rank degradation.
}
\label{tab:retriever_eq_rankdrop}
\end{table*}

%

\section{Full Dataset-by-Model-by-Form Retrieval Results}
\label{app:full_3way_retrieval}

Section~\ref{sec:query_form_sensitivity} reports marginal summaries along the
dataset form and dataset model axes.
Those summaries are compact, but they average over one axis of variation.
For completeness, Table~\ref{tab:appendix_full_dataset_model_form_results} reports the
full dataset,model,query-form retrieval breakdown.
The table follows the main outcome flow and includes only aggregate Recall@5
and paired MeanRD.

\begin{table*}[t]
\centering
\tiny
\setlength{\tabcolsep}{2.5pt}
\renewcommand{\arraystretch}{0.86}
\begin{adjustbox}{max width=.78\textwidth}
\begin{tabular}{llrrrr}
\toprule
\textbf{Model} &
\textbf{Query form} &
\textbf{Source R@5} &
\textbf{CORA R@5} &
$\boldsymbol{\Delta}$\textbf{R@5} &
\textbf{MeanRD} \\
\midrule
\multicolumn{6}{@{}l}{\textbf{AudioCaps}} \\
\midrule
LAION
& Statement  & 54.1 & 40.2 & -13.9 & 32.16 \\
& Question   & 54.1 & 38.4 & -15.7 & 62.50 \\
& Key phrase & 54.1 & 37.9 & -16.2 & 58.76 \\
& Command    & 54.1 & 38.2 & -15.9 & 47.29 \\
& Indirect   & 54.1 & 41.8 & -12.3 & 47.43 \\
\addlinespace[2pt]

M2D
& Statement  & 68.7 & 56.7 & -12.0 & 21.63 \\
& Question   & 68.7 & 59.3 & -9.4  & 13.53 \\
& Key phrase & 68.7 & 64.8 & -3.9  & 0.24 \\
& Command    & 68.7 & 43.5 & -25.2 & 59.77 \\
& Indirect   & 68.7 & 53.8 & -15.0 & 18.79 \\
\addlinespace[2pt]

MGA
& Statement  & 63.2 & 61.7 & -1.6  & 4.28 \\
& Question   & 63.2 & 55.6 & -7.7  & 19.69 \\
& Key phrase & 63.2 & 61.7 & -1.6  & 2.62 \\
& Command    & 63.2 & 59.4 & -3.8  & 13.13 \\
& Indirect   & 63.2 & 35.9 & -27.3 & 115.80 \\
\addlinespace[2pt]

MS-CLAP
& Statement  & 10.1 & 37.0 & 26.9 & -143.23 \\
& Question   & 10.1 & 39.0 & 29.0 & -151.06 \\
& Key phrase & 10.1 & 9.0  & -1.0 & 32.72 \\
& Command    & 10.1 & 39.0 & 29.0 & -152.71 \\
& Indirect   & 10.1 & 34.1 & 24.0 & -135.51 \\

\midrule
\bottomrule
\end{tabular}
\end{adjustbox}
\caption{{
Full dataset--model--query-form retrieval results under CORA reformulation (AudioCaps). R@5 and $\Delta$R@5 are percentage points; MeanRD is $r_{\mathrm{CORA}}-r_{\mathrm{src}}$.
}}
\label{tab:appendix_full_dataset_model_form_results}
\end{table*}

\begin{table*}[t]
\centering
\tiny
\setlength{\tabcolsep}{2.5pt}
\renewcommand{\arraystretch}{0.86}
\begin{adjustbox}{max width=.78\textwidth}
\begin{tabular}{llrrrr}
\toprule
\textbf{Model} &
\textbf{Query form} &
\textbf{Source R@5} &
\textbf{CORA R@5} &
$\boldsymbol{\Delta}$\textbf{R@5} &
\textbf{MeanRD} \\
\midrule
\multicolumn{6}{@{}l}{\textbf{Clotho}} \\
\midrule
LAION
& Statement  & 29.9 & 32.9 & 3.0   & -14.36 \\
& Question   & 29.9 & 25.6 & -4.3  & 2.69 \\
& Key phrase & 29.9 & 13.9 & -16.0 & 164.85 \\
& Command    & 29.9 & 27.7 & -2.2  & -13.62 \\
& Indirect   & 29.9 & 23.3 & -6.5  & 20.05 \\
\addlinespace[2pt]

M2D
& Statement  & 37.9 & 38.1 & 0.2   & -1.75 \\
& Question   & 37.9 & 32.8 & -5.1  & 5.22 \\
& Key phrase & 37.9 & 21.1 & -16.8 & 49.86 \\
& Command    & 37.9 & 36.2 & -1.7  & 1.70 \\
& Indirect   & 37.9 & 27.8 & -10.2 & 29.95 \\
\addlinespace[2pt]

MGA
& Statement  & 37.8 & 37.0 & -0.8 & 0.10 \\
& Question   & 37.8 & 35.9 & -1.9 & 18.08 \\
& Key phrase & 37.8 & 36.0 & -1.8 & 5.61 \\
& Command    & 37.8 & 35.6 & -2.2 & 17.78 \\
& Indirect   & 37.8 & 33.8 & -4.0 & 40.26 \\
\addlinespace[2pt]

MS-CLAP
& Statement  & 34.0 & 35.3 & 1.3   & -9.94 \\
& Question   & 34.0 & 30.9 & -3.1  & 2.06 \\
& Key phrase & 34.0 & 7.7  & -26.3 & 239.36 \\
& Command    & 34.0 & 35.1 & 1.1   & -10.13 \\
& Indirect   & 34.0 & 30.5 & -3.5  & 6.51 \\

\midrule
\bottomrule
\end{tabular}
\end{adjustbox}
\caption{{
Full dataset--model--query-form retrieval results under CORA reformulation (Clotho). R@5 and $\Delta$R@5 are percentage points; MeanRD is $r_{\mathrm{CORA}}-r_{\mathrm{src}}$.
}}

\end{table*}

\begin{table*}[t]
\centering
\tiny
\setlength{\tabcolsep}{2.5pt}
\renewcommand{\arraystretch}{0.86}
\begin{adjustbox}{max width=.78\textwidth}
\begin{tabular}{llrrrr}
\toprule
\textbf{Model} &
\textbf{Query form} &
\textbf{Source R@5} &
\textbf{CORA R@5} &
$\boldsymbol{\Delta}$\textbf{R@5} &
\textbf{MeanRD} \\
\midrule
\multicolumn{6}{@{}l}{\textbf{MACS}} \\
\midrule
LAION
& Statement  & 16.2 & 0.8  & -15.4 & 361.91 \\
& Question   & 16.2 & 4.2  & -12.0 & 299.02 \\
& Key phrase & 16.2 & 16.2 & 0.0   & 2.15 \\
& Command    & 16.2 & 8.8  & -7.4  & 166.96 \\
& Indirect   & 16.2 & 10.4 & -5.8  & 243.89 \\
\addlinespace[2pt]

M2D
& Statement  & 1.4 & 5.8 & 4.4  & -169.65 \\
& Question   & 1.4 & 3.4 & 2.0  & -141.18 \\
& Key phrase & 1.4 & 1.6 & 0.2  & 50.01 \\
& Command    & 1.4 & 7.8 & 6.4  & -206.73 \\
& Indirect   & 1.4 & 1.2 & -0.2 & 64.83 \\
\addlinespace[2pt]

MGA
& Statement  & 8.6 & 8.0  & -0.6 & 5.72 \\
& Question   & 8.6 & 14.2 & 5.6  & -34.27 \\
& Key phrase & 8.6 & 9.0  & 0.4  & 5.56 \\
& Command    & 8.6 & 7.6  & -1.0 & 5.67 \\
& Indirect   & 8.6 & 14.2 & 5.6  & -23.43 \\
\addlinespace[2pt]

MS-CLAP
& Statement  & 14.0 & 4.6  & -9.4 & 110.13 \\
& Question   & 14.0 & 9.2  & -4.8 & 63.43 \\
& Key phrase & 14.0 & 12.4 & -1.6 & 22.37 \\
& Command    & 14.0 & 6.4  & -7.6 & 92.76 \\
& Indirect   & 14.0 & 5.4  & -8.6 & 74.67 \\

\midrule
\bottomrule
\end{tabular}
\end{adjustbox}
\caption{{
Full dataset--model--query-form retrieval results under CORA reformulation (MACS). R@5 and $\Delta$R@5 are percentage points; MeanRD is $r_{\mathrm{CORA}}-r_{\mathrm{src}}$.
}}

\end{table*}

\begin{table*}[t]
\centering
\tiny
\setlength{\tabcolsep}{2.5pt}
\renewcommand{\arraystretch}{0.86}
\begin{adjustbox}{max width=.78\textwidth}
\begin{tabular}{llrrrr}
\toprule
\textbf{Model} &
\textbf{Query form} &
\textbf{Source R@5} &
\textbf{CORA R@5} &
$\boldsymbol{\Delta}$\textbf{R@5} &
\textbf{MeanRD} \\
\midrule
\multicolumn{6}{@{}l}{\textbf{MECAT}} \\
\midrule
LAION
& Statement  & 9.7 & 7.0  & -2.7 & 87.86 \\
& Question   & 9.7 & 10.4 & 0.7  & 161.31 \\
& Key phrase & 9.7 & 3.8  & -5.9 & 323.46 \\
& Command    & 9.7 & 9.9  & 0.2  & 132.71 \\
& Indirect   & 9.7 & 9.4  & -0.2 & 233.13 \\
\addlinespace[2pt]

M2D
& Statement  & 12.6 & 12.7 & 0.1  & 8.76 \\
& Question   & 12.6 & 14.0 & 1.4  & 5.68 \\
& Key phrase & 12.6 & 15.9 & 3.3  & -113.27 \\
& Command    & 12.6 & 9.3  & -3.3 & 166.85 \\
& Indirect   & 12.6 & 13.9 & 1.3  & -11.71 \\
\addlinespace[2pt]

MGA
& Statement  & 13.3 & 14.4 & 1.1  & -0.51 \\
& Question   & 13.3 & 12.5 & -0.8 & 93.64 \\
& Key phrase & 13.3 & 14.6 & 1.3  & 18.39 \\
& Command    & 13.3 & 12.4 & -0.9 & 73.97 \\
& Indirect   & 13.3 & 11.9 & -1.4 & 157.28 \\
\addlinespace[2pt]

MS-CLAP
& Statement  & 8.1 & 5.3  & -2.8 & -53.57 \\
& Question   & 8.1 & 2.7  & -5.4 & 143.58 \\
& Key phrase & 8.1 & 19.3 & 11.2 & -386.79 \\
& Command    & 8.1 & 3.2  & -5.0 & 251.05 \\
& Indirect   & 8.1 & 2.1  & -6.0 & 136.80 \\
\bottomrule
\end{tabular}
\end{adjustbox}
\caption{{
Full dataset--model--query-form retrieval results under CORA reformulation (MECAT). R@5 and $\Delta$R@5 are percentage points; MeanRD is $r_{\mathrm{CORA}}-r_{\mathrm{src}}$.
}}

\end{table*}

The full breakdown reinforces the main conclusion that CORA sensitivity is interactional.
A query form that is stable for one dataset--model pair can be harmful for
another, and aggregate $\Delta$R@5 should be interpreted together with paired
MeanRD.

\section{Interaction Analysis for Query-Form Sensitivity}
\label{sec:query_interaction_prediction}

Section~\ref{sec:query_form_sensitivity} shows that CORA sensitivity varies
along both dataset--form and dataset--model axes. We further test whether this
pattern is better described by interactional effects rather than by additive
model, dataset, and query-form effects alone.

We predict rank degradation using grouped cross-validation by audio ID. This
split prevents CORA pairs derived from the same target audio from
appearing in both the training and validation folds. The prediction target is
$\mathrm{RD}(p)$, where positive values indicate that the target rank becomes
worse after reformulation.

Table~\ref{tab:query_interaction_prediction} reports prediction performance
under increasingly specific feature sets. Additive model, dataset, and
query-form terms provide limited predictive power. Adding model-dataset and
query interaction terms improves performance, and the full
model-dataset-query cell gives the highest correlation. This supports the
main-text conclusion that query-form sensitivity is conditional on the
model-dataset regime rather than being a universal property of the surface
query form.

\begin{table}[t]
\centering
\small
\setlength{\tabcolsep}{4pt}
\begin{tabular}{lrr}
\toprule
\textbf{Feature terms} & \textbf{Pearson $r$} & \textbf{Spearman $\rho$} \\
\midrule
additive model/dataset/query & 0.240 & 0.244 \\
model-dataset + query & 0.285 & 0.257 \\
pairwise query interactions & 0.309 & 0.317 \\
full model-dataset-query cell & 0.376 & 0.391 \\
\bottomrule
\end{tabular}
\caption{
Grouped cross-validated prediction of RD. Interaction terms improve prediction
over additive model, dataset, and query form terms, indicating that CORA
sensitivity is conditional on the model-dataset regime.
}
\label{tab:query_interaction_prediction}
\end{table}


\section{Hard Negative Definition Sensitivity}
\label{app:hn_sensitivity}

Section~\ref{sec:embedding_diagnostic_metrics} defines TBMD using the strongest
non-target audio for each query. Specifically, the source query uses
$h_{\mathrm{src}}$, the top-1 non-target audio under $q_{\mathrm{src}}$, and
the CORA query uses $h_{\mathrm{cora}}$, the top-1 non-target audio under
$q_{\mathrm{cora}}$. This dynamic top-1 definition is the main diagnostic used
in the paper:
\begin{equation}
\small
\begin{aligned}
\mathrm{TBMD}_{\mathrm{dyn}}
&= \big[\cos(q_{\mathrm{src}}, a_{\mathrm{tgt}})
      - \cos(q_{\mathrm{src}}, h_{\mathrm{src}})\big] \\
&\quad - \big[\cos(q_{\mathrm{cora}}, a_{\mathrm{tgt}})
      - \cos(q_{\mathrm{cora}}, h_{\mathrm{cora}})\big].
\end{aligned}
\end{equation}

We include a sensitivity check to verify that the boundary signal is not an
artifact of this particular hard negative anchoring choice. We compare the main
dynamic top-1 definition against two fixed-top-1 alternatives. The first fixes
the hard negative to the strongest CORA-query non-target audio and evaluates
both query forms against that same competitor:
\begin{equation}
\small
\begin{aligned}
\mathrm{TBMD}_{\mathrm{CORA\text{-}HN}}
&= \big[\cos(q_{\mathrm{src}}, a_{\mathrm{tgt}})
      - \cos(q_{\mathrm{src}}, h_{\mathrm{cora}})\big] \\
&\quad - \big[\cos(q_{\mathrm{cora}}, a_{\mathrm{tgt}})
      - \cos(q_{\mathrm{cora}}, h_{\mathrm{cora}})\big].
\end{aligned}
\end{equation}
The second fixes the hard negative to the strongest source-query non-target
audio and evaluates both query forms against that source-side competitor:
\begin{equation}
\small
\begin{aligned}
\mathrm{TBMD}_{\mathrm{src\text{-}HN}}
&= \big[\cos(q_{\mathrm{src}}, a_{\mathrm{tgt}})
      - \cos(q_{\mathrm{src}}, h_{\mathrm{src}})\big] \\
&\quad - \big[\cos(q_{\mathrm{cora}}, a_{\mathrm{tgt}})
      - \cos(q_{\mathrm{cora}}, h_{\mathrm{src}})\big].
\end{aligned}
\end{equation}
All three variants are degradation-oriented, so larger values indicate a larger
loss of target margin after reformulation.

\begin{table}[t]
\centering
\small
\setlength{\tabcolsep}{6pt}
\begin{tabular}{lr}
\toprule
\textbf{Top-1 boundary degradation definition} & \textbf{Pearson w/ RD} \\
\midrule
CORA-HN fixed TBMD & 0.487 \\
Source-HN fixed TBMD & 0.317 \\
Dynamic top-1 TBMD (pooled) & \textbf{0.598} \\
\bottomrule
\end{tabular}
\caption{
Sensitivity of TBMD to top-1 hard negative anchoring choices.
All variants use top-1 non-target audios, matching the main diagnostic setting.
The dynamic top-1 definition used in the main text gives the strongest
association with RD.
}
\label{tab:top1_hn_definition_sensitivity}
\end{table}

Table~\ref{tab:top1_hn_definition_sensitivity} shows that all top-1 boundary
variants are positively associated with RD. The fixed-source variant is
weaker because a source-side hard negative can miss a new competitor that
becomes salient only after reformulation. The fixed-CORA variant is stronger,
but it evaluates the source and CORA queries against a boundary selected only
after reformulation. The main dynamic top-1 definition is strongest because it
compares each query against its own most competitive non-target audio. This
supports the use of dynamic top-1 TBMD as the primary boundary diagnostic.

\section{Anchor and Content Controls for Target Alignment Loss}
\label{app:anchor_content_robustness}

The main diagnostic analysis uses source queries as anchors. This choice is
central to the CORA protocol: the source query defines the original benchmark
retrieval input, and CORA measures how the same target audio changes rank when
that query is reformulated into user style CORA queries. However, this design
also raises a possible concern. The observed Target Alignment Loss (TAL) may be
tied to the particular source-query anchor, or to lexical and content-realization
differences between source queries and CORA queries. We therefore include two
supporting controls that modify the anchoring condition.

First, in the statement-anchor control, we use the Statement query as the anchor
and compare the remaining four CORA query forms against it. We denote this
setting as CORA4. Because the Statement query is itself a CORA query and is
closest in form to a descriptive retrieval query, this control reduces
dependence on the original source query while preserving a paired
reformulation setting. Second, in the content-fixed wrapper control, the core
acoustic content is held fixed and only the surrounding query wrapper is varied.
This control tests whether the TAL signal persists when the main variation
comes from query formulation rather than changes in the described sound content.

These controls are not intended to replace the main source-query-anchored
analysis, because the primary CORA setting is precisely the shift from source
queries to user style CORA queries. Instead, they serve as robustness checks for
whether the TAL--RD association is unique to the source-query comparison.

\begin{table}[t]
\centering
\small
\setlength{\tabcolsep}{5pt}
\begin{tabular}{lrr}
\toprule
\textbf{Anchor/control} & \textbf{N} & $\mathbf{r}(\textbf{TAL}, \textbf{RD})$ \\
\midrule
Source query $\rightarrow$ CORA query      & 66,100 & 0.615 \\
Statement query $\rightarrow$ CORA4        & 45,888 & 0.399 \\
Content-fixed wrapper control              & 32,320 & 0.406 \\
\bottomrule
\end{tabular}
\caption{
Anchor and content controls for Target Alignment Loss (TAL), computed with the pooled cross-dataset candidate set.
TAL is
degradation-oriented, so positive correlations indicate that larger loss of
target alignment is associated with larger RD.
}
\label{tab:anchor_content_robustness}
\end{table}

Table~\ref{tab:anchor_content_robustness} shows that the TAL--RD
association becomes weaker when the original source-query anchor is removed or
when the acoustic content is fixed. This reduction is expected, since these
controls no longer exactly match the main CORA protocol. Nevertheless, the
association remains consistently positive. The result suggests that the
target-alignment diagnostic is not solely an artifact of using source queries
as anchors, nor solely a consequence of source-query-to-CORA-query content
mismatch. We therefore treat these results as supporting robustness checks for
the main source-query-anchored analysis.

\section{TBMD Control Analysis}
\label{app:tbmd_control}

Because TBMD is computed from the same similarity scores that determine rank, its correlation with RD in Section~\ref{sec:main_diagnostic} could partly reflect this construction rather than a genuine boundary effect. We test whether TBMD adds explanatory power over TAL once we control for dataset, retrieval model, query form, source rank, and candidate pool size.

For both the per-dataset and pooled cross-dataset settings (66,100 pairs each), we fit a 5-fold grouped cross-validated Ridge regression (Appendix~\ref{sec:proj_regression}) predicting RD, first from TAL alone and then from TAL together with dynamic-HN TBMD, both with fixed effects for dataset, model, and query form. We also compute the partial correlation of TBMD with RD after residualizing out TAL, dataset, model, query form, $\Delta_{\mathrm{move}}$, source rank, and pool size via OLS.

\begin{table}[t]
\centering
\begin{tabular}{lcc}
\toprule
\textbf{Analysis} & \textbf{Per-dataset} & \textbf{Pooled} \\
\midrule
TAL-only (Pearson $r$) & 0.561 & 0.615 \\
TAL + dynamic-HN TBMD & 0.644 & 0.626 \\
Increment & +0.083 & +0.010 \\
Partial $r$ beyond TAL & 0.382 & 0.144 \\
\bottomrule
\end{tabular}
\caption{Control analysis for TBMD. Adding dynamic-HN TBMD to TAL improves RD prediction in both settings, and TBMD keeps a positive partial correlation with RD once TAL, dataset, model, query form, $\Delta_{\mathrm{move}}$, source rank, and pool size are held fixed.}
\label{tab:tbmd_control}
\end{table}

TBMD improves prediction over TAL alone in both settings, and its partial correlation with RD stays positive once these confounds are held fixed, though the partial association is smaller in the pooled setting. We therefore treat TBMD as a diagnostic signal that adds information beyond TAL, rather than evidence that TBMD alone drives rank degradation.

\section{Boundary-Margin Token Saliency Audit}
\label{sec:token_saliency_appendix}

We further examine whether the boundary-centered RD pattern is visible at the token-attribution level.
This analysis is intended as a local diagnostic audit rather than a causal token-deletion experiment.
For each unchanged CORA query, we compute token-level Gradient $\times$ Input saliency with respect to the target boundary margin.

We evaluate the four CLAP retrievers used in the main experiments: LAION-CLAP, MS-CLAP, M2D-CLAP, and MGA-CLAP.
The analysis is restricted to sentence-like CORA forms: Statement, Question, Command, and Indirect.
We exclude Key phrase queries because compressed phrase queries do not share the same function-word and intent-word structure as the other query forms.
For each model, we sample up to 10 examples per dataset, query form, and RD group, yielding 320 query pairs per model and 1,280 query pairs in total.
The two RD groups are:
\[
\text{rank-worsened}: r_{\mathrm{cora}} > r_{\mathrm{src}},
\]
\[
\text{rank-improved/stable}: r_{\mathrm{cora}} \leq r_{\mathrm{src}}.
\]

For a CORA query $q$, let $a_{\mathrm{tgt}}$ be the target audio and let $a_{\mathrm{hn}}$ be the highest-scoring non-target audio under the CORA query.
The local boundary margin is
\[
M(q) = \cos(q,a_{\mathrm{tgt}}) - \cos(q,a_{\mathrm{hn}}).
\]
A positive margin means that the query scores the target above the local hard negative, while a negative margin means that the hard negative is favored over the target.
For each token $i$, with input embedding $e_i$, we compute Gradient $\times$ Input saliency with respect to this margin:
\[
s_i = e_i \cdot \frac{\partial M(q)}{\partial e_i}.
\]
We normalize token saliency within each query by total absolute saliency:
\[
\tilde{s}_i = \frac{s_i}{\sum_j |s_j| + \epsilon}.
\]
Positive saliency indicates local support for the target over the hard negative, while negative saliency indicates local support for the hard negative over the target.

Tokens are grouped into three surface categories:
\begin{itemize}
    \item \textbf{acoustic content}: tokens describing sound events, sources, acoustic attributes, or scenes;
    \item \textbf{query intent}: tokens such as \textit{find}, \textit{search}, \textit{retrieve}, \textit{audio}, and \textit{sound};
    \item \textbf{function/form}: determiners, auxiliaries, prepositions, question markers, politeness markers, and other form-related tokens.
\end{itemize}
For each group, we separately aggregate positive and negative saliency mass.
For example, Content+ measures positive boundary-margin support assigned to acoustic-content tokens, while Form- measures negative boundary-margin contribution assigned to function/form tokens.

\begin{table*}[t]
\centering
\small
\setlength{\tabcolsep}{5pt}
\resizebox{\textwidth}{!}{%
\begin{tabular}{llrrrrrrr}
\toprule
\textbf{Model} & \textbf{RankDrop group}
& \textbf{N}
& \textbf{MeanRD}
& \textbf{Content+}
& \textbf{Content-}
& \textbf{Form-}
& \textbf{Intent-}
& \textbf{Mean margin} \\
\midrule
LAION   & worsened        & 160 &  71.819 & 0.2832 & 0.2387 & 0.1205 & 0.0521 & -0.2189 \\
LAION   & improved/stable & 160 & -37.863 & 0.2799 & 0.2647 & 0.1143 & 0.0409 & -0.0928 \\
MS-CLAP & worsened        & 160 &  45.581 & 0.2774 & 0.2840 & 0.1025 & 0.0486 & -0.1250 \\
MS-CLAP & improved/stable & 160 & -50.213 & 0.2696 & 0.2885 & 0.1191 & 0.0389 & -0.0697 \\
M2D     & worsened        & 160 &  31.356 & 0.2780 & 0.2915 & 0.1178 & 0.0404 & -0.0430 \\
M2D     & improved/stable & 160 & -23.938 & 0.3025 & 0.2710 & 0.1243 & 0.0466 & -0.0169 \\
MGA     & worsened        & 160 &  36.269 & 0.3080 & 0.3267 & 0.1009 & 0.0420 & -0.1138 \\
MGA     & improved/stable & 160 & -19.813 & 0.3166 & 0.3231 & 0.0946 & 0.0452 & -0.0459 \\
\bottomrule
\end{tabular}
}
\caption{
Token-group saliency shares by RD group.
Mean margin is the local target boundary margin under the CORA query.
Across all four models, rank-worsened pairs have lower mean boundary margins than rank-improved or stable pairs.
}
\label{tab:rankdrop_token_saliency_groups}
\end{table*}

Table~\ref{tab:rankdrop_token_saliency_groups} shows that the clearest group-level separation is the boundary margin itself.
Across all four models, rank-worsened pairs have lower mean target boundary margins than rank-improved or stable pairs.
For LAION, the mean margin decreases from $-0.0928$ to $-0.2189$; for MS-CLAP, from $-0.0697$ to $-0.1250$; for M2D, from $-0.0169$ to $-0.0430$; and for MGA, from $-0.0459$ to $-0.1138$.
This is consistent with the main TBMD result: harmful reformulations are associated with weaker local target-boundary margins.

However, the token-group shares themselves do not provide a stable explanation of RD.
Content+, Content-, Form-, and Intent- vary only modestly between the two RD groups, and no token category separates rank-worsened pairs consistently across all models.
For example, Form- is slightly higher for rank-worsened pairs in LAION and MGA, but lower in MS-CLAP and M2D.
This suggests that RD is not reducible to a single harmful token class such as function words or intent words.

We next test whether token-group saliency features improve RD prediction beyond the boundary diagnostics used in the main text.
We use grouped cross-validation by audio ID and predict continuous RD.
The feature sets are: fixed effects only, query movement, token-group saliency, boundary features, and boundary features plus token-group saliency.
Boundary features include query movement, target-alignment loss, TBMD, and boundary projection.

\begin{table}[t]
\centering
\small
\setlength{\tabcolsep}{4pt}
\begin{tabular}{lrr}
\toprule
\textbf{Feature set} & \textbf{Pearson $r$} & \textbf{Spearman $\rho$} \\
\midrule
Fixed effects only & 0.077 & 0.007 \\
Query movement & 0.115 & 0.045 \\
Token-group saliency & 0.090 & 0.037 \\
Boundary features & 0.634 & 0.591 \\
Boundary + token saliency & 0.632 & 0.581 \\
\bottomrule
\end{tabular}
\caption{
Grouped cross-validated RD prediction from token saliency and boundary features.
Token-group saliency alone has weak predictive value, and adding it to boundary features does not improve prediction.
}
\label{tab:rankdrop_token_saliency_prediction}
\end{table}

Table~\ref{tab:rankdrop_token_saliency_prediction} shows that token-group saliency alone is weak for continuous RD prediction, reaching only Pearson $r=0.090$ and Spearman $\rho=0.037$ in the pooled setting.
Query movement is also weak, with Pearson $r=0.115$ and Spearman $\rho=0.045$.
By contrast, boundary features reach Pearson $r=0.634$ and Spearman $\rho=0.591$.
Adding token-group saliency to the boundary features does not improve prediction; performance changes from $r=0.634$, $\rho=0.591$ to $r=0.632$, $\rho=0.581$.

This audit gives a limited but useful role to token-level attribution.
Token saliency confirms that rank-worsened queries have poorer local target boundary margins, supporting the boundary-centered interpretation.
At the same time, coarse token groups do not explain RD beyond the boundary diagnostics.
We therefore treat token saliency as a descriptive local attribution tool rather than evidence that a particular token class causally drives CORA degradation.

\section{Dataset-Level Diagnostic Correlations}
\label{app:dataset_level_diagnostic}

Section~\ref{sec:tbmd_main_analysis} reports the pooled diagnostic correlations
between RD and the three embedding-space signals. Here, we provide the
dataset-level breakdown to verify that the main pattern is not driven by a
single dataset. For each dataset, we compute Pearson correlations between
RD and query movement, Target Alignment Loss (TAL), and Target Boundary
Margin Degradation (TBMD).

\begin{table}[t]
\centering
\small
\setlength{\tabcolsep}{4pt}
\resizebox{\columnwidth}{!}{%
\begin{tabular}{lrrrrrr}
\toprule
\textbf{Dataset} & \textbf{N} & \textbf{MeanRD} & \textbf{Worsened (\%)} 
& $\mathbf{r(\Delta_{\mathrm{move}})}$ & $\mathbf{r(\mathrm{TAL})}$ & $\mathbf{r(\mathrm{TBMD})}$ \\
\midrule
AudioCaps & 19,260 & -1.61 & 43.2 & -0.016 & 0.531 & 0.547 \\
Clotho    & 19,880 & 27.71 & 51.6 & 0.221  & 0.580 & 0.585 \\
MACS      & 10,000 & 49.69 & 50.2 & 0.361  & 0.737 & 0.716 \\
MECAT     & 16,960 & 71.43 & 55.5 & 0.216  & 0.700 & 0.668 \\
\bottomrule
\end{tabular}%
}

\caption{
Dataset-level diagnostic correlations with RD. Worsened denotes the
percentage of pairs for which the CORA-query rank is lower than the source-query
rank. TAL and TBMD are consistently more associated with RD than query
movement across datasets.
}
\label{tab:dataset_level_diagnostic}
\end{table}

Table~\ref{tab:dataset_level_diagnostic} shows that the main diagnostic pattern
is not specific to a single dataset. Query movement has weak or variable
association with RD, ranging from near zero on AudioCaps to moderate
association on MACS. In contrast, TAL and TBMD remain positively associated
with RD across all four datasets. This supports the interpretation that
CORA degradation is better characterized by target-alignment loss and
target-boundary margin degradation than by raw movement away from the source
query.

\section{Projection Analysis Details}
\label{app:projection_analysis_details}

This appendix provides the formal details for the projection analysis in the
main text. The goal of this analysis is to test whether CORA-induced RD
is better explained by the \emph{direction} of the source-to-CORA query shift
than by the shift magnitude alone. All embeddings are $\ell_2$-normalized, and
cosine similarity is therefore computed as an inner product. For any non-zero
vector $x$, we write
\[
\widehat{x}=\frac{x}{\|x\|_2+\varepsilon}
\]
for numerical stability.

\subsection{Mathematical Formulation}
\label{sec:proj_math}

Given a source caption embedding $q_{\mathrm{src}}$ and the corresponding CORA
query embedding $q_{\mathrm{cora}}$, we first compute the raw source-to-CORA
shift:
\[
\delta = q_{\mathrm{cora}} - q_{\mathrm{src}} .
\]
Because the retrieval embeddings lie on the unit sphere, we remove the radial
component of this shift and analyze the tangent-space displacement at the source
caption anchor:
\[
d
=
\delta - \langle q_{\mathrm{src}}, \delta \rangle q_{\mathrm{src}} .
\]
The movement-magnitude feature used in the main projection experiment is
\[
\|d\|_2 .
\]
We intentionally denote this feature as $\|d\|_2$, rather than
$\Delta_{\mathrm{move}}$, to distinguish it from the cosine-distance query
movement metric used in the TBMD diagnostic section.

\paragraph{Target-audio projection.}
To measure whether the query shift moves toward the target audio, we first
project the target audio embedding onto the same tangent space:
\[
u_{\mathrm{audio}}
=
\widehat{
 a_{\mathrm{tgt}}
 - \langle q_{\mathrm{src}}, a_{\mathrm{tgt}} \rangle q_{\mathrm{src}}
} .
\]
The target-audio projection is then
\[
p_{\mathrm{audio}}
=
\langle d, u_{\mathrm{audio}} \rangle .
\]
A larger positive value indicates that the CORA shift moves in a direction more
aligned with the target audio representation, whereas a negative value indicates
movement away from the target direction.

\paragraph{Boundary projection.}
To measure whether the query shift preserves local separation between the
target and a nearby non-target competitor, we define a target 
boundary direction. Given the source-anchored hard negative $h_{\mathrm{src}}$
defined in Appendix~\ref{sec:proj_hn}, the tangent-space boundary direction is
\[
u_{\mathrm{boundary}}
=
\widehat{
(a_{\mathrm{tgt}} - h_{\mathrm{src}})
-
\langle q_{\mathrm{src}}, a_{\mathrm{tgt}} - h_{\mathrm{src}} \rangle
q_{\mathrm{src}}
} .
\]
The boundary projection is
\[
p_{\mathrm{boundary}}
=
\langle d, u_{\mathrm{boundary}} \rangle .
\]
This feature approximates whether the source-to-CORA query shift moves in a
direction that increases or decreases the local target-minus-hard-negative score
difference. Positive values indicate movement in a target-favoring boundary
direction, while negative values indicate movement toward weakening the local
retrieval boundary.

\subsection{Hard Negative Selection}
\label{sec:proj_hn}

The projection experiment uses a source-anchored hard negative so that the
analysis tests whether the CORA shift preserves the original local retrieval
boundary. This choice is complementary to the dynamic hard negative definition
used for TBMD: TBMD measures realized margin degradation under each query,
whereas the projection analysis measures the direction of the source-to-CORA
shift relative to a source-side boundary.

For each audio item $i$ and CORA form $f$, we first construct a boundary
candidate pool from both retrieval neighborhoods. Let $S_{\mathrm{src}}^5(i)$
be the top-5 non-target audios retrieved by the source query, and let
$S_f^5(i)$ be the top-5 non-target audios retrieved by the CORA query of form
$f$. The target audio is excluded from both sets. We define
\[
P_{i,f}
=
S_{\mathrm{src}}^5(i) \cup S_f^5(i) .
\]
The final hard negative is selected as the highest-scoring non-target audio
under the source query within this union pool:
\[
h_{\mathrm{src}}^{(i,f)}
=
\arg\max_{a_j \in P_{i,f}}
\cos(q_{\mathrm{src}}^{(i)}, a_j) .
\]
Using the union pool makes the candidate boundary sensitive to competitors that
are salient either before or after reformulation, while selecting the final hard
negative with the source query keeps the analyzed boundary anchored to the
pre-reformulation retrieval state.

\subsection{Regression Details}
\label{sec:proj_regression}

We conduct the projection analysis on AudioCaps, Clotho, MACS, and MECAT using
four CLAP-family retrievers: LAION-CLAP, M2D-CLAP, MGA-CLAP, and MS-CLAP. For
each source caption, we evaluate five CORA query forms: Key phrase, Statement,
Question, Command, and Indirect. The final evaluation contains 66,100 CORA
pairs.

The prediction target is continuous RD,
\[
\mathrm{RD}
=
\mathrm{rank}_{\mathrm{cora}} - \mathrm{rank}_{\mathrm{src}},
\]
where larger values indicate that the target audio is ranked lower after
reformulation. We use 5-fold grouped cross-validation with groups defined by
\texttt{dataset::audio\_id}, preventing different CORA forms from the same audio
sample from appearing in both the training and test folds.

\begin{flushleft}
The regression pipeline consists of
\texttt{DictVectorizer},
\texttt{StandardScaler} with \texttt{with\_mean=False}, and
\texttt{Ridge} with \texttt{alpha=1.0} and \texttt{solver=lsqr}.
\end{flushleft} All settings include fixed effects for
dataset, retrieval model, and query form. We then compare the following feature
sets:
\begin{itemize}
    \item fixed effects only;
    \item fixed effects + $\|d\|_2$;
    \item fixed effects + $p_{\mathrm{audio}}$;
    \item fixed effects + $p_{\mathrm{boundary}}$.
\end{itemize}
Each projection row adds only the named continuous feature to the same
fixed-effect baseline. This isolates whether RD is better explained by
movement magnitude, target-oriented direction, or boundary-oriented direction.
Prediction quality is measured by Pearson correlation and Spearman correlation
between out-of-fold predictions and ground-truth RD values.

\subsection{Interpretation}
\label{sec:proj_interpretation}

The projection analysis is designed to complement the scalar TAL and TBMD
diagnostics in the main text. TAL and TBMD measure realized changes in target
alignment and target boundary margin. By contrast, the projection
features ask whether the actual embedding displacement from the source caption
to the CORA query points in a target-preserving or boundary-preserving
direction.

The main results show that adding $\|d\|_2$ gives only a small improvement over
fixed effects, while $p_{\mathrm{audio}}$ and $p_{\mathrm{boundary}}$ provide
substantially stronger RD prediction. This supports the directional
boundary hypothesis: harmful reformulations are not simply those that move far
from the source caption, but those whose embedding shift fails to move toward
the target audio or fails to preserve the local target
boundary.

\section{Additional Inference-Time Fusion Check}
\label{sec:fusion_check}

We additionally evaluate simple inference-time score fusion baselines that average scores from multiple CLAP retrievers.
These baselines are included as supporting checks for aggregate retrieval strength, not as robustness-oriented training methods.
As shown in Table~\ref{tab:fusion_check}, score fusion improves aggregate Source R@5 and CORA R@5 over individual CLAP retrievers, but it is not treated as evidence of query-form robustness.

\begin{table}[t]
\centering
\small
\setlength{\tabcolsep}{4pt}
\resizebox{\columnwidth}{!}{%
\begin{tabular}{lrrrr}
\toprule
\textbf{Retriever}
& \textbf{Source R@5}
& \textbf{CORA R@5}
& $\boldsymbol{\Delta}$\textbf{R@5}
& \textbf{MeanRD} \\
\midrule
M2D & 34.89 & 29.57 & -5.31 &  2.46 \\
MGA & 34.52 & 31.69 & -2.83 & 30.34 \\
MultiCLAP-ScoreAvg-Z & 38.67 & 35.75 & -2.92 & 24.29 \\
MultiCLAP-ScoreAvg-MinMax & 38.28 & 35.50 & -2.77 & 25.04 \\
\bottomrule
\end{tabular}%
}

\caption{
Additional inference-time score fusion check under the CORA protocol.
M2D and MGA are included as strong single-retriever references.
R@5 and $\Delta$R@5 values are reported as percentages.
}
\label{tab:fusion_check}
\end{table}

\section{Query Preprocessing}
\label{sec:preprocessing}

We examine whether query preprocessing can recover retrieval performance degraded by query reformulation and whether the diagnostic pattern observed in our main analysis persists after preprocessing. Our main experiments evaluate the retrievers directly, without applying any external query rewriting. In contrast, the experiments in this section introduce an external preprocessing module before retrieval (preprocessor $\rightarrow$ retriever).

We evaluate three preprocessing conditions on a balanced sample of 500 source items, with 125 items sampled from each of the four datasets used in our main experiments (Clotho, AudioCaps, MECAT, and MACS). We use all five CORA query forms and four CLAP retrievers, with retrieval performed over the full candidate pool of each dataset.

We consider three conditions: \textbf{No canonicalizer}, \textbf{General rewriting}, and \textbf{CORA-aware rewriting}. In the No canonicalizer condition, each CORA query is directly provided to the retriever. General rewriting uses a generic rewriter that is not given the CORA query-form taxonomy. CORA-aware rewriting, in contrast, is given the query's CORA form and the definition of that form, allowing the rewriter to use this information when rewriting the query into a caption-style description. However, such form information is not generally available in real-world settings. Therefore, we treat CORA-aware rewriting as a type-aware diagnostic condition rather than a directly deployable preprocessing method. The prompts used for both rewriting conditions are provided in Appendix~\ref{app:preprocessing-prompts}.

\begin{table}[t]
\centering
\small
\setlength{\tabcolsep}{3pt}
\begin{tabular*}{\columnwidth}{@{\extracolsep{\fill}}lccccc@{}}
\toprule
Condition & RD & R@5 & $\Delta$move & TAL & TBMD \\
\midrule
No canonicalizer
    & 6.33 & 37.33 & 0.103 & 0.551 & 0.653 \\
General rewriting
    & 4.14 & 38.10 & 0.080 & 0.519 & 0.634 \\
CORA-aware
    & -0.03 & 40.02 & 0.020 & 0.546 & 0.683 \\
\bottomrule
\end{tabular*}
\caption{
Effect of query preprocessing on retrieval performance and diagnostic
associations. RD denotes Mean RankDrop, and R@5 is reported in percent.
$\Delta$move, TAL, and TBMD report Pearson's $r$ with RD.
Source-query R@5 is 38.95\% in all conditions.
}
\label{tab:preprocessing}
\end{table}

As shown in Table~\ref{tab:preprocessing}, query preprocessing reduces average rank degradation. With 95\% cluster-bootstrap CIs, General and CORA-aware rewriting reduce MeanRD by 2.18 (CI [0.27, 4.15]) and 6.36 ranks (CI [4.86, 7.91]). In particular, CORA-aware rewriting reduces MeanRD from +6.33 to $-0.03$ while increasing R@5 from 37.33\% to 40.02\%.

However, this improvement is not uniform across individual items or query forms. CORA-aware rewriting improves item-level MeanRD for 377 of the 500 items, while 118 items worsen and 5 remain unchanged. At the query-form level, Key Phrase is the only exception in MeanRD, increasing from +3.03 to +3.78, although its R@5 still improves.

Despite these changes in retrieval performance, the relative ordering of the diagnostic associations remains consistent across all three conditions. Raw query movement remains only weakly associated with RD, whereas TAL and TBMD show stronger associations. For example, under CORA-aware rewriting, $r(\Delta_{\mathrm{move}}, \mathrm{RD})=0.020$, compared with $r(\mathrm{TAL}, \mathrm{RD})=0.546$ and $r(\mathrm{TBMD}, \mathrm{RD})=0.683$. These results show that query rewriting can improve retrieval performance degraded by reformulation at the pipeline level, while the TAL/TBMD and movement diagnostic pattern persists after preprocessing.

\section{Prompt}
\subsection{Query Prompt Definitions}
\label{tab:querydefinition}

\begin{tcolorbox}[systemprompt, title=Key Phrase Prompt]
\textbf{System Prompt}\\
You are an expert at optimizing audio captions for search engines.
Analyze the given caption and compress it into a tight Key Phrase.
 
\textbf{Constraints}
\begin{enumerate}[leftmargin=*, nosep, topsep=4pt]
  \item Strictly NO complete sentences (subject + verb). You must use a noun phrase format.
  \item Remove unnecessary prepositions and conjunctions.
  \item Convert long background descriptions into concise modifiers.
  \item Preserve the texture and action of the sound using present participles.
  \item Avoid conjunction-heavy structures.
\end{enumerate}
 
\medskip
\textbf{User Prompt}\\
Source Caption: \texttt{\{caption\}}\\[4pt]
Key phrase:
\end{tcolorbox}
 
 
\begin{tcolorbox}[systemprompt, title=Statement Prompt]
\textbf{System Prompt}\\
You are an expert at objectively describing audio scenarios.
Analyze the given caption and write a full descriptive statement.
 
\textbf{Constraints}
\begin{enumerate}[leftmargin=*, nosep, topsep=4pt]
  \item Generate a grammatically complete sentence.
  \item Convert stiff noun phrases into natural verbal structures.
  \item Preserve simultaneous/background events using conjunctions.
  \item End the sentence with a period.
\end{enumerate}
 
\medskip
\textbf{User Prompt}\\
Source Caption: \texttt{\{caption\}}\\[4pt]
Statement:
\end{tcolorbox}
 
 
\begin{tcolorbox}[systemprompt, title=Question Prompt]
\textbf{System Prompt}\\
You are an expert at generating questions to verify audio content.
Analyze the given caption and create a natural yes/no question.
 
\textbf{Constraints}
\begin{enumerate}[leftmargin=*, nosep, topsep=4pt]
  \item Preserve important audible content.
  \item Convert the caption into a fluent yes/no question.
  \item Avoid overly mechanical wording.
  \item End the query with a question mark.
\end{enumerate}
 
\medskip
\textbf{User Prompt}\\
Source Caption: \texttt{\{caption\}}\\[4pt]
Question:
\end{tcolorbox}
 
 
\begin{tcolorbox}[systemprompt, title=Command Prompt]
\textbf{System Prompt}\\
You are an expert at crafting precise instructions for search systems or agents.
Analyze the given caption and generate a direct command.
 
\textbf{Constraints}
\begin{enumerate}[leftmargin=*, nosep, topsep=4pt]
  \item Start directly with an action verb.
  \item Use concise search-oriented directives.
  \item Preserve important audible content.
  \item End the query with a period.
\end{enumerate}
 
\medskip
\textbf{User Prompt}\\
Source Caption: \texttt{\{caption\}}\\[4pt]
Command:
\end{tcolorbox}
 
 
\begin{tcolorbox}[systemprompt, title=Indirect Prompt]
\textbf{System Prompt}\\
You are an expert in highly polite and conversational communication.
Create an indirect or polite request asking to find the sounds described in the caption.
 
\textbf{Constraints}
\begin{enumerate}[leftmargin=*, nosep, topsep=4pt]
  \item Begin with a natural polite expression.
  \item Smoothly connect the request to the sound retrieval task.
  \item Preserve important audible content.
  \item Use appropriate punctuation depending on sentence form.
\end{enumerate}
 
\medskip
\textbf{User Prompt}\\
Source Caption: \texttt{\{caption\}}\\[4pt]
Indirect:
\end{tcolorbox}
 
 
\subsection*{Output Schema}
 
For \textit{question}, \textit{command}, and \textit{indirect}, the model additionally follows:
 
\begin{tcolorbox}[outputschema, title=Output Schema]
Return only a valid JSON object with exactly these string fields:
\begin{itemize}[leftmargin=*, nosep, topsep=4pt]
  \item \texttt{"answer"}: the final query text.
  \item \texttt{"explanation"}: a concise explanation of how you chose the final query.
\end{itemize}
 
\medskip
In \texttt{"explanation"}, first generate $\{\texttt{N}\}$ plausible starting words or phrases for the final query,
then choose a final starting word or phrase that is \emph{not} in that generated list.
The \texttt{"answer"} must start with that non-listed choice.
Explain briefly that the final query starts with the non-listed choice.
\end{tcolorbox}

\subsection{Real-User Query Classification Prompt}
\label{app:wildclaims-prompt}

\begin{tcolorbox}[systemprompt, title=Real-User Query Classification Prompt]
\textbf{System Prompt}\\
You are an expert evaluator for classifying user queries by linguistic form.

Your task is to classify the given user question into exactly one of the following six labels:
\begin{enumerate}[leftmargin=*, nosep, topsep=4pt]
    \item Key Phrase
    \item Statement
    \item Question
    \item Command
    \item Indirect
    \item None
\end{enumerate}

Classify based only on the linguistic form and user intent of the text.
Do not judge factuality.

\medskip
\textbf{Query Type Guidelines}

\medskip
\textbf{1. Key Phrase}\\
A compressed noun-phrase-style query.

\textit{Characteristics:}
\begin{itemize}[leftmargin=*, nosep, topsep=2pt]
    \item Not a complete sentence.
    \item No explicit subject-verb sentence structure.
    \item Uses concise phrase composition.
    \item Often omits function words.
\end{itemize}

\textit{Example:} ``Rain pattering on metal with distant thunder''

\medskip
\textbf{2. Statement}\\
A declarative sentence.

\textit{Characteristics:}
\begin{itemize}[leftmargin=*, nosep, topsep=2pt]
    \item Complete sentence with subject and verb.
    \item States or describes something directly.
    \item Usually ends with a period.
\end{itemize}

\textit{Example:} ``There is the sound of rain pattering on metal while thunder rumbles in the distance.''

\medskip
\textbf{3. Question}\\
A direct question asking for information or whether something is true.

\textit{Characteristics:}
\begin{itemize}[leftmargin=*, nosep, topsep=2pt]
    \item Uses interrogative form or question wording.
    \item Often starts with words such as ``What'', ``Why'', ``How'', ``Can'', ``Does'', ``Is'', ``Are'', or ``Who''.
    \item Usually ends with a question mark.
\end{itemize}

\textit{Example:} ``Is there a recording of rain pattering on metal while thunder rumbles in the distance?''

\medskip
\textbf{4. Command}\\
A direct instruction or imperative request.

\textit{Characteristics:}
\begin{itemize}[leftmargin=*, nosep, topsep=2pt]
    \item Starts with an imperative verb such as ``Find'', ``Search'', ``Locate'', ``Retrieve'', ``Tell'', ``Explain'', or ``List''.
    \item No explicit subject is needed.
    \item Directly tells a system or person to do something.
\end{itemize}

\textit{Example:} ``Search for a recording of rain pattering on metal while thunder rumbles in the distance.''

\medskip
\textbf{5. Indirect}\\
A polite, softened, or indirect request.

\textit{Characteristics:}
\begin{itemize}[leftmargin=*, nosep, topsep=2pt]
    \item Uses expressions such as ``Could you please'', ``I would appreciate it if you could'', ``I was wondering if you might'', or ``Would it be possible''.
    \item Requests an action without using a direct imperative command.
\end{itemize}

\textit{Example:} ``Could you please locate an audio file featuring rain pattering on metal while thunder rumbles in the distance?''

\medskip
\textbf{6. None}\\
Use this label only when the text does not fit any of the five labels above.

\textit{Examples:}
\begin{itemize}[leftmargin=*, nosep, topsep=2pt]
    \item Empty or nonsensical text.
    \item A fragment that is not a meaningful key phrase.
    \item Metadata, markup, or text that is not a user query.
\end{itemize}

\medskip
\textbf{User Prompt}\\
User Question: \texttt{\{query\}}\\[4pt]
Query Type:
\end{tcolorbox}

\subsection{Query Preprocessing Prompts}
\label{app:preprocessing-prompts}

\begin{tcolorbox}[systemprompt, title=General Rewriting Prompt]
\textbf{System Prompt}\\
You are an expert at rewriting user queries for the audio retrieval process.

\medskip
\textbf{User Prompt}\\
Rewrite the following audio search query to improve information retrieval, preserving the original intent and all audible content while improving clarity and adding only relevant acoustic context. Return only the rewritten query.

\medskip
Input Query: \texttt{\{query\}}\\
Output:
\end{tcolorbox}


\begin{tcolorbox}[systemprompt, title=CORA-Aware Rewriting Prompt]
\textbf{System Prompt}\\
\textbf{Instruction:} You are an expert at rewriting user queries for the audio retrieval process.

\medskip
\textbf{Requirement:} Given an audio search query and its query type, your task is to provide a rewritten caption-style query. Please refer to the query rewriting guidelines and examples.

\medskip
\textbf{Query Types and Rewriting Guidelines}

\medskip
\textbf{Question:} Convert a yes/no or interrogative query into a direct declarative description of the audible content. Remove expressions such as ``Can you hear,'' ``Does this audio contain,'' and ``Is there a recording of.''

\medskip
\textbf{Command:} Convert a direct retrieval instruction into a description of the audible content. Remove imperative and retrieval-oriented expressions such as ``Find,'' ``Search for,'' ``Locate,'' and ``Retrieve.''

\medskip
\textbf{Indirect:} Convert a polite or softened retrieval request into a direct description of the audible content. Remove politeness markers, conversational wrappers, hedging, and indirect request expressions.

\medskip
\textbf{Key Phrase:} Preserve the compressed phrase structure while rewriting it as a concise and natural audio caption.

\medskip
\textbf{Statement:} Preserve the directly described sound scene while removing redundant wording and making the description concise.

\medskip
\textbf{Examples}

\medskip
\textbf{Input Query:} Is there a recording of rain pattering on metal while thunder rumbles in the distance?\\
\textbf{Action:} Question\\
\textbf{Output:} [``Rain pattering on metal while thunder rumbles in the distance.'']

\medskip
\textbf{Input Query:} Search for a recording of glass breaking followed by people shouting.\\
\textbf{Action:} Command\\
\textbf{Output:} [``Glass breaking followed by people shouting.'']

\medskip
\textbf{Input Query:} Could you please locate an audio file featuring waves crashing against rocks?\\
\textbf{Action:} Indirect\\
\textbf{Output:} [``Waves crashing against rocks.'']

\medskip
\textbf{Input Query:} Birds chirping, light wind, distant traffic\\
\textbf{Action:} Key Phrase\\
\textbf{Output:} [``Birds chirping with light wind and distant traffic.'']

\medskip
\textbf{Input Query:} A train passes while a warning bell rings.\\
\textbf{Action:} Statement\\
\textbf{Output:} [``A train passing while a warning bell rings.'']

\medskip
\textbf{User Prompt}\\
Input Query: \texttt{\{query\}}\\
Action: \texttt{\{query\_type\}}\\
Output:
\end{tcolorbox}

\end{document}